*Research Article*

# Dynamical System Perspective of Cosmological Models Minimally Coupled with Scalar Field

**S. Surendra Singh and Chingtham Sonia**

*Department of Mathematics, NIT Manipur, Imphal 795004, India*

Correspondence should be addressed to Chingtham Sonia; chingtham.sonia19@gmail.com

The stability criteria for the dynamical system of a homogeneous and isotropic cosmological model are investigated with the interaction of a scalar field in the presence of a perfect fluid. In this paper, we depict the dynamical system perspective to study qualitatively the scalar field cosmology under two special cases, with and without potential. In the absence of potential, we get a two-dimensional dynamical system, and we study the analytical as well as geometrical behavior. For the dynamical system with potential, we analyze different potential forms: simple exponential potential form ($V(\phi) = V_o e^{-\lambda\phi}$), double exponential potential form $V(\phi) = V_o \exp(-A \exp(\sqrt{2}\alpha\phi))$, and inverse power law potential form ($V(\phi) = V_o \phi^{-\alpha}$). We generate an autonomous system of ordinary differential equations (ASODE) for each case by introducing new dimensionless variables and obtain respective fixed points. We also analyze the type, nature, and stability of the fixed points and how their behavior reflects towards the cosmological scenarios. Throughout the whole work, the investigation of this model has shown us the deep connection between these theories and cosmic acceleration phenomena. The phase plots of the system at different conditions and different values of $\gamma$ have been analyzed in detail, and their geometrical interpretations have been studied. The perturbation plots of the dynamical system have been analyzed with emphasis on our analytical findings. We have evaluated the total energy density ($\Omega_\phi$) at the fixed points and also found out the suitable range of $\gamma$ and $\lambda$ for a stable model.

## 1. Introduction

The dynamical system approach, nowadays, has become one of the most suitable and viable ways for qualitative specification of various cosmic features, specially for studying the possible asymptotic states at early as well as late times of the evolving Universe. We do a qualitative study of the system rather than finding the exact solutions. By qualitative study, we mean the study of the behavior, obtaining information about the properties of the system. Considering the early Universe, inflation should be taken into account with which the expansion of the Universe with acceleration can be expounded through the dominance of potential energy ($V(\phi)$) of a scalar field ($\phi$) over its energy density. In fact, inflation has been regarded as a part of cosmological evolution. In recent years, it is well discovered that our Universe is in the phase of not just expansion but expansion with acceleration [1]. In terms of observational evidence, we can cite observations from supernova light curve data to Wilkinson Microwave Anisotropy Probe (WMAP) data [2] which is in agreement with the current phase of the accelerating Universe. We can also mention here the $f(R)$ theory of gravity for explaining the expanding Universe [3]. Several intricate observations have admitted that our Universe restarted expansion with acceleration around 5 Gyr ago and at low energy scales of approximately $10^{-4}$ eV [2, 4, 5]. But it is quite an accepted fact that inflation is one such theory of exponential expansion of the early Universe which corresponds to energy scales $\sim 10^{16}$ GeV [6–13]. This large scale deviation in the energy scales have drawn many researchers towards these two phenomena and how the difference has come up. This curiosity has led researchers to hunt new cosmic models that best suit the observations. So far, among all the recent works, the approaches that give appealing descriptions on the present scenario of the expanding Universe are firstly, "dark energy" (DE) associated with a large negative pressure which is frequently described by a conventional vacuum energy or scalar field, and secondly, we can mention the six-parameter base $\Lambda$CDM model which also agrees with the current high-precision data [1, 14] though there are some cosmological constant and coincidence problems [15–20]. Also, the Dvali-Gabadadze-Porrati (DGP) braneworld model [21–23] has put forward the cosmic accelerated expansion of the present Universe. We can also name the modified theories of gravity in this context [24–26]. Despite roaming about all these cosmological journeys, in this paper, we will only stick to the dynamical system perspective to cosmology. A cosmological model described by an autonomous system of ordinary differential equation (ASODE) having a past time attractor, one or two saddle points, and a late time attractor is regarded as the most suitable and complete model. The past time attractor represents the inflationary epoch, saddle points correspond to the phase of the radiation- and matter-dominated Universe, and the late time attractor represents the accelerated expansion phase of the present Universe [27, 28]. Scalar field cosmology with scalar potential forms such as flat, constant, simple exponential potential, inverse power law potential, double-exponential potential, and PNGB potential has been studied by many researchers also [29–39]. Scalar field cosmology has also paved its grand way to study dark energy which is taken as the main entity for the cosmic expansion with acceleration at late time. Indeed, scalar field cosmological models have owned their esteemed place in modern cosmology, specially in analyzing early inflation as well as late time acceleration. Dark energy scenarios for the scalar field are such an exotic matter that provokes the necessary negative pressure so as to cause acceleration [40]. One can use a relativistic equation of state like that of radiation ($\gamma = 4/3$) or an ultrarelativistic fluid ($\gamma = 2$). One can also choose $\gamma = 0$ to fetch a vacuum energy density with no actual fluid where we obtain $\rho = -p$. So, it is worthy enough to confer about the situation with a perfect fluid having the above equation of state for various values of $\gamma$ such that $0 \leq \gamma \leq 2$.

This paper has been arranged as follows: in Section 2, we depict how the system of gravitational field and wave equations are developed. We introduce new dimensionless variables in these field equations to generate a dynamical system. We again categorize this section into two subsections: one for analysis without potential and the other for analysis with potential. From the perspective of a dynamical system which is an autonomous system of ordinary differential equation (ASODE), we are going to analyze each subsection deeply by investigating the fixed points for each case. We study the nature of stability of the possible fixed points obtained in respective cases and discuss their stability along with the cosmic scenarios. We devote Section 3 to the conclusion.

## 2. Dynamical System Analysis

A dynamical system is a mathematical system that describes the time dependence of the position of a point in the space that surrounds it, termed as ambient space. There are different ways of approach to a dynamical system, namely, the measure theory approach which is motivated by the ergodic theory, real dynamical system, discrete dynamical system, etc. Here, we are approaching the system through an autonomous system of ordinary differential equations (ASODE). ASODE is a system of ordinary differential equations which does not depend explicitly on time. As for our dynamical system, we will be using logarithmic time ($N = \ln(a/a_0)$) as our independent variable, where $a$ represents the scale factor and $a_0$ denotes the present scale factor value which, later on, will be taken to be unity for simplicity as doing so does not affect the behavior of our system. Now, we consider a minimally coupled classical scalar field ($\phi$) represented by the classical equation of motion,

$$\ddot{\phi} + 3H\dot{\phi} + \frac{dV}{d\phi} = 0, \tag{1}$$

where the overhead dot represents the derivative with respect to cosmic time "$t$." Here, in this paper, we consider that the Universe is filled with a perfect fluid having an equation of state as follows:

$$p = (\gamma - 1)\rho, \tag{2}$$

where $\gamma$ is a constant, $p$ is the pressure, and $\rho$ is the density of the fluid. We also consider the Universe to be characterized by the following line element for an FRW space time as

$$ds^2 = -dt^2 + a^2(t)\left[\frac{dt^2}{1-kr^2} + r^2 d\theta^2 + r^2 \sin^2\theta d\phi\right]. \tag{3}$$

Here, $a(t)$ stands for a scale factor, and $k$ is the curvature parameter which can take values −1, 0, or +1, according to open, flat, or close Universe, respectively. We assume the energy momentum tensor of the perfect fluid to be characterized by the following equation:

$$T_{\mu\nu}^{(F)} = (\rho + p)u_\mu v_\nu + p g_{\mu\nu}, \tag{4}$$

where $p$ and $\rho$ are, respectively, thermodynamic pressure and density of the fluid. The unit time vector $v_\mu$ for a comoving system is specified by $v^\mu v_\mu = -1$ and $v^\mu = \delta_o^\mu$. We consider a minimally coupled classical scalar field $\phi$ which contributes to $T_{\mu\nu} = T_{\mu\nu}^{(F)} + T_{\mu\nu}^{(\phi)}$ as follows:

$$T_{\mu\nu}^{(\phi)} = \phi_{,\mu}\phi_{,\nu} - g_{\mu\nu}\left[\frac{1}{2}\phi, \alpha\phi^{,\alpha}\right] + V(\phi). \tag{5}$$

Equations (4) and (5) lead to the gravitational field equations as the following:

$$3H^2 + 3\frac{k}{a^2} = \frac{1}{2}\dot{\phi}^2 + V + \rho, \tag{6}$$

$$2\dot{H} + 3H^2 + \frac{k}{a^2} = -\frac{1}{2}\dot{\phi}^2 + V - \rho, \tag{7}$$

Table 1: Table for analysis without potential.

| Fixed points | $x$ | $y$ | Eigenvalues | Stable range of $\gamma$ | $\Omega_\phi$ | Behavior |
|---|---|---|---|---|---|---|
| $P$ | 0 | 0 | $-4, (2-3\gamma)$ | Stable for $\gamma \in (2/3, 2]$ | 0 | Stable for $2/3 < \gamma \leq 2$ |
| $Q$ | 0 | 1 | $\left(-3\gamma - \frac{16}{3}, (3\gamma-2)\right)$ | Stable for $0 \leq \gamma < 2/3$ | 1 | Late time attractor for $0 \leq \gamma < 2/3$ |
| $R$ | 1 | 0 | $(6-3\gamma), 4$ | Unstable for all $\gamma$ | 0 | Simple repelling node, unstable |
| $S$ | 1 | $\frac{3(\gamma-2)}{(3\gamma-2)}$ | $\pm\sqrt{(\gamma-0.72)(\gamma+7.58)}$ | Unstable for all $\gamma$ | $\frac{3(\gamma-2)}{(3\gamma-2)}$ | Unstable |

where the overhead dot denotes differentiation w.r.t. cosmic time $t$ and $H = \dot{a}/a$ is the Hubble parameter. Here, we choose the units in such a way that $8\pi G = 1$ and $c = 1$ [41]. The conservation equation of the fluid is described by

$$\dot{\rho} + 3H(\rho + p) = 0. \tag{8}$$

*2.1. Analysis without Potential.* For studying the cosmological model without potential, we take $V(\phi) = 0$ and using equations (1) and (4)–(7), we have the following field and wave equations:

$$\begin{aligned} \ddot{\phi} + 3H\dot{\phi} &= 0, \\ T_{\mu\nu}^{(\phi)} &= \phi_{,\mu}\phi_{,\nu} - g_{\mu\nu}\left[\frac{1}{2}\phi_{,\alpha}\phi^{,\alpha}\right], \\ 3H^2 + 3\frac{k}{a^2} &= \frac{1}{2}\dot{\phi}^2 + \rho, \\ 2\dot{H} + 3H^2 + \frac{k}{a^2} &= -\frac{1}{2}\dot{\phi}^2 - p. \end{aligned} \tag{9}$$

We want to develop a dynamical model in the presence of the curvature constant ($k$). We are developing a generalized system in the presence of the curvature constant ($k$). To analyze the dynamical system in a flat Universe, the curvature constant ($k$) is taken to be zero, and with the introduction of suitable normalized variables, the system can be studied [42]. To create a dynamical system out of the above field and wave equations, we present a new set of dimensionless variables as $x = \dot{\phi}^2/6H^2$, $y = \rho/3H^2$, and $N = \ln\ a$. Using these new variables, the above equations reduce to the following set of autonomous system of ordinary differential equations (ASODE):

$$\begin{aligned} x' &= 4x^2 + 3\left(\gamma - \frac{2}{3}\right)xy - 4x, \\ y' &= 3\left(\gamma - \frac{2}{3}\right)y^2 + 4yx + (2-3\gamma)y, \end{aligned} \tag{10}$$

where the overhead dash represents differentiation w.r.t. logarithmic time, $N = \ln\ a$, and curvature parameter, $k = H^2a^2(x + y - 1)$. The total relative energy density of the homogeneous scalar field is given by

$$\Omega_\phi = \frac{k^2\rho\phi}{3H^2}, \tag{11}$$

where $\rho_\phi$ denotes the energy density of the scalar field. To find the fixed points, we equate $x' = 0$ and $y' = 0$. We have got four fixed points as $P(0,0)$, $Q(0,1)$, $R(1,0)$, and $S(1, 3(\gamma-2)/(3\gamma-2))$ and check the stability of the fixed points so obtained. For this, we determine the Jacobian matrix ($J$) of the autonomous system at the respective fixed points. We see that the $y$ coordinate of the fixed point $S$ becomes infinite when $\gamma = 2/3$.

So, in order to study all the fixed points in the finite phase plane, we have to analyze the system taking $\gamma \neq 2/3$. The fixed points along with the associated eigenvalues are shown in Table 1. We have also analyzed the range value of $\gamma$ within which we got the stable model. We have also calculated the value of $\Omega_\phi$ at each fixed point. The Jacobian matrix ($J_P$) derived at the fixed point $P(0,0)$ is given by

$$J_P = \begin{bmatrix} -4 & 0 \\ 0 & (2-3\gamma) \end{bmatrix}. \tag{12}$$

Since $J_P$ is a diagonal matrix, the eigenvalues are given by the diagonal entries which are $-4$ and $(2-3\gamma)$. Both the eigenvalues are real and will have negative signs for $\gamma > 2/3$. Since we have taken $\gamma \neq 2/3$, the fixed point $P$ remains a hyperbolic fixed point as none of the eigenvalues vanish. Here, stability depends on the sign of the eigenvalues of the Jacobian matrix ($J_P$). Since both the eigenvalues becomes negative when $\gamma > 2/3$, the fixed point $P$ is stable and a simple attracting node for those values for $\gamma > 2/3$. For $\gamma < 2/3$, the fixed point $P$ acts as a saddle point since it has one positive real eigenvalue and one negative real eigenvalue. We have calculated the value of $\Omega_\phi$ at the fixed point $P$ which is obtained as $\Omega_\phi = 0$. From another point of view, we have calculated the perturbation along the $x$- and $y$-axes for $P$ from the dynamical system as $\epsilon_x = C_1\ \exp(-4N)$ and $\epsilon_y = C_2\ \exp((2-3\gamma)N)$, where $C_1$ and $C_2$ are arbitrary constants. Figures 1 and 2 show the perturbation plots along the $x$- and $y$-axes with respect to $N$, respectively. We see that the projection of perturbation along the $x$-axis ($\epsilon_x$) decays to zero with the increase in $N$. For the values of $\gamma > 2/3$, the perturbation along $y$-axis ($\epsilon_y$) monotonically decreases to zero as $N$ tends to infinity. Also from the phase plot shown in Figures 3 and 4, it is observed that all the trajectories close enough to $P$ move towards it for $\gamma > 2/3$. This shows the stability of fixed point $P$ for all $\gamma > 2/3$ which is also in agreement with the geometrical approach. With these analytical and geometrical findings at this fixed point $P$, we obtain a barotropic fluid-dominated solution with $\gamma > 2/3$ and $\Omega_\phi = 0$. As the upper limit of $\gamma$ for any type of fluid is 2 [43], so the fixed point $P$ is stable for ultrarelativistic fluid ($\gamma = 2$), matter dominated ($\gamma = 1$), and radiation cosmology ($\gamma = 4/3$). For $\gamma < 2/3$, we see that some trajectories are attracted towards the point $P$ while some trajectories deviate away from $P$. On the other hand, for $\gamma < 2/3$, the fixed point $P$ becomes a saddle which means an unstable condition. The Jacobian matrices at the fixed points $Q(0, 1)$, $R(1, 0)$, and $S(1, 3(\gamma - 2)/(3\gamma - 2))$ are, respectively, given by

$$J_Q = \begin{bmatrix} \left(-3\gamma - \frac{16}{3}\right) & 0 \\ 4 & (3\gamma - 2) \end{bmatrix},$$
$$J_R = \begin{bmatrix} 4 & -3\left(\gamma - \frac{2}{3}\right) \\ 0 & (-3\gamma + 6) \end{bmatrix}, \tag{13}$$
$$J_S = \begin{bmatrix} 4 - 3(\gamma - 2) & (-3\gamma + 2) \\ \frac{12(\gamma - 2)}{(3\gamma - 2)} & (3\gamma - 6) \end{bmatrix}.$$

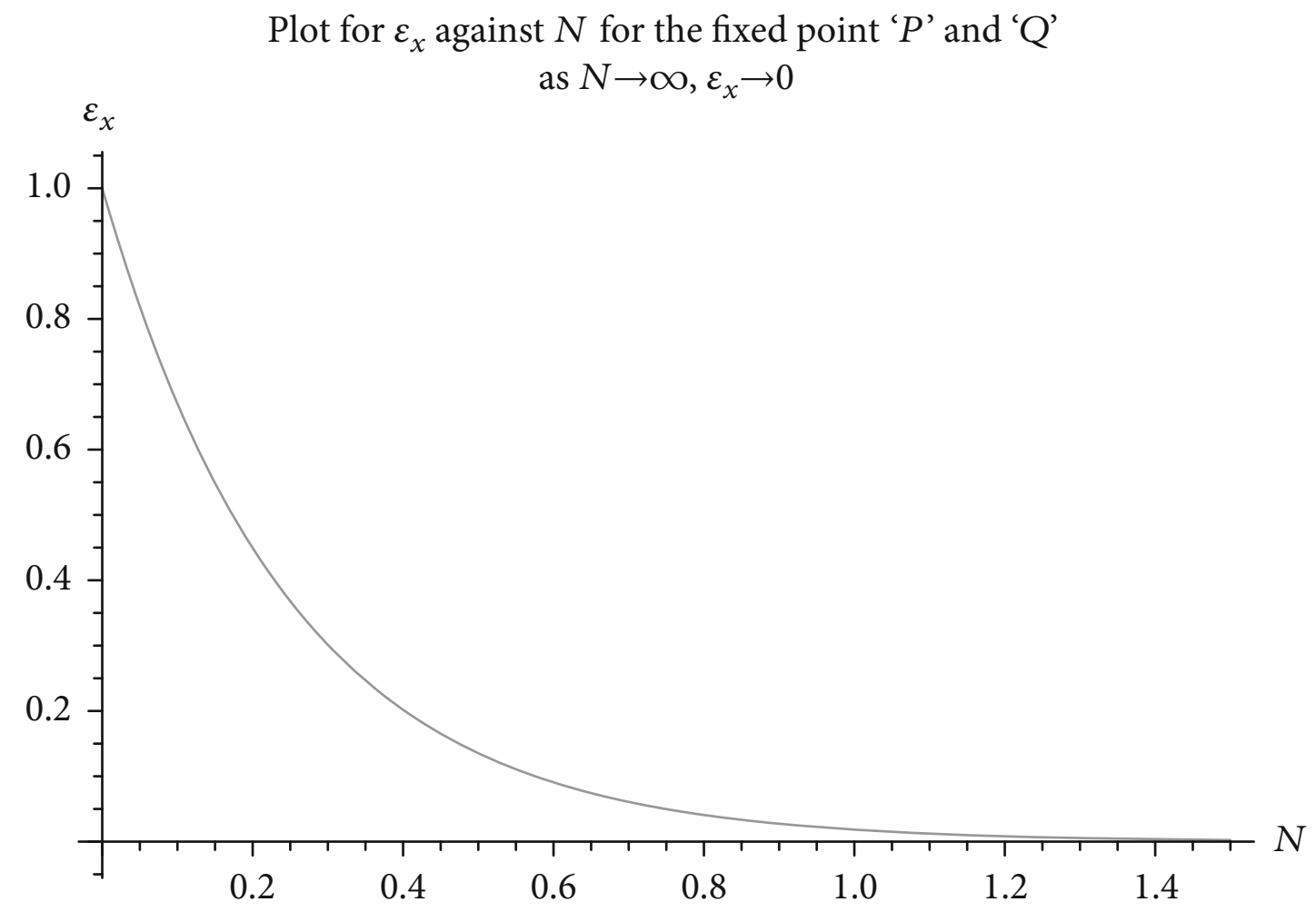

Figure 1: Shows the variation of perturbations along the $x$-axis ($\epsilon_x$) against $N$ for fixed points $P$ and $Q$.

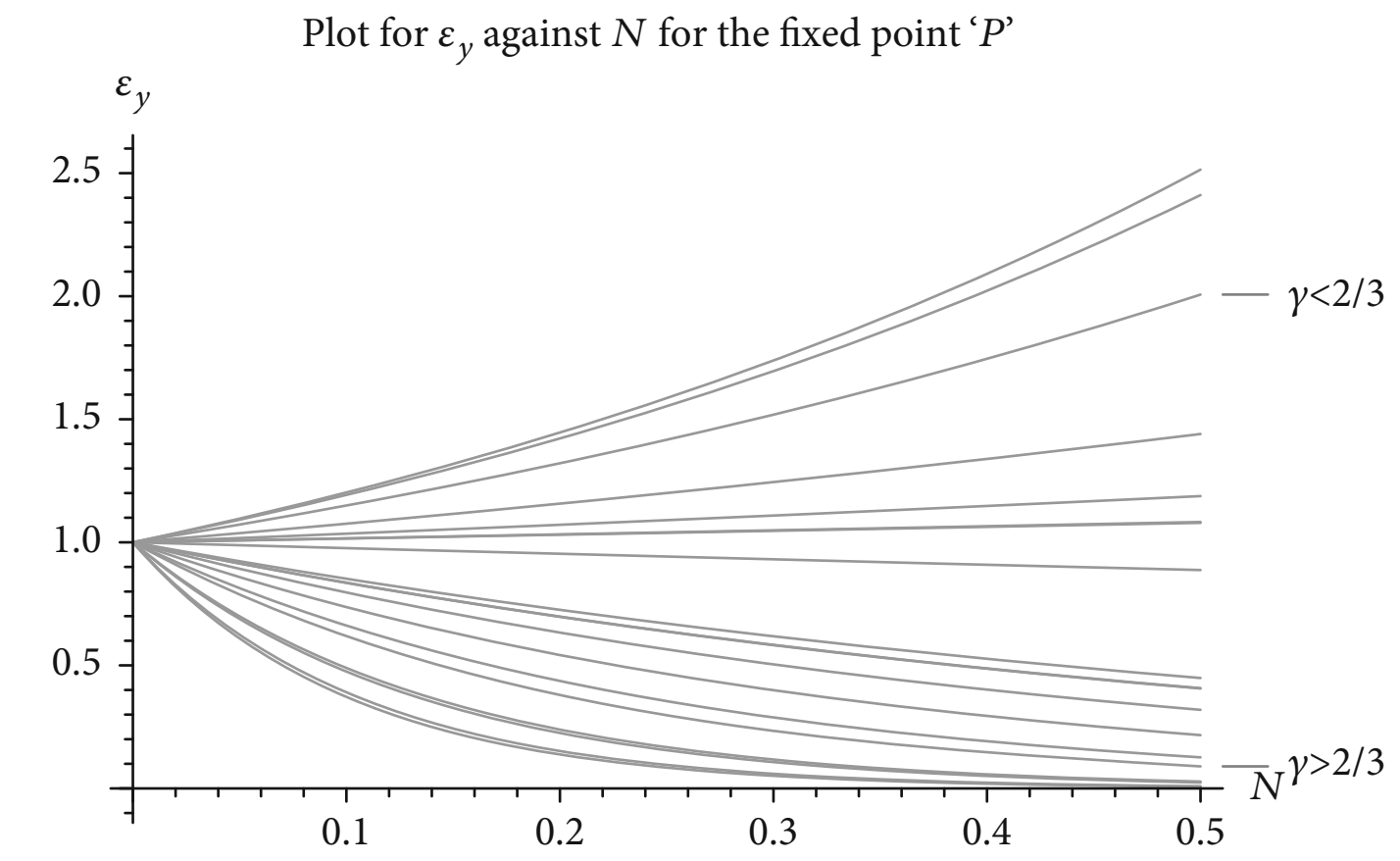

Figure 2: Shows the variation of perturbations along the $y$-axis ($\epsilon_y$) at $\gamma > 2/3$ and $\gamma < 2/3$ for fixed point $P$.

The eigenvalues of the Jacobian matrix at the respective fixed points and their stability analysis are shown in Table 1.

For the fixed point $Q$, both eigenvalues become negative for $\gamma < 2/3$. So the fixed point $Q$ is stable for all values of $\gamma$ such that $0 \le \gamma \le 2/3$. We have also calculated the value of $\Omega_\phi$ at $Q$ which is found to be $\Omega_\phi = 1$. We find the perturbation along the $x$-axis ($\epsilon_x$) as a function of $N$, $\epsilon_x = D_1 \exp(-N)$ where $D_1$ is an arbitrary constant. This is an exponentially decreasing function of $N$. So when $N$ tends to infinity, $\epsilon_x$ decays to zero. The perturbation along the $y$-axis ($\epsilon_y$) is found to be $\epsilon_y = D_2 \exp(-(2 - 3\gamma)N)$. When $\gamma < 2/3$, $\epsilon_y$ monotonically decreases with the increase of $N$. We also plot the perturbation along the $y$-axis ($\epsilon_y$) against $N$ at $Q$ which is shown in Figure 5. From this figure, we see that as $N$ tends to infinity, $\epsilon_y$ monotonically decreases to zero. As the perturbation fails to grow with the increase of $N$, the fixed point $Q$ is stable for all those values of $\gamma$ satisfying $0 \le \gamma \le 2/3$. From the phase plots shown in Figures 6 and 7, we see that all the curves near $Q$ are attracted towards it for $\gamma = 0$, but if we take $\gamma$ to be 4/3, all the nearby trajectoriesclose enough to $Q$ are being deviated away from $Q$. This shows that $Q$ is unstable. Thus, the analytic as well as geometric inferences reflect that this scalar-dominated solution with $\Omega_\phi = 1$ is a late time attractor for $0 \leq \gamma < 2/3$. So the fixed point $Q$ is stable for $\gamma = 0$ where pressure becomes equal in magnitude and opposite in sign to the energy density. The stable fixed point $Q$ assures the presence of this negative pressure which gives a dark energy model. For the fixed points $R$ and $S$, the perturbation along the $x$-axis for both the fixed points comes out to be a monotonically increasing function of $N$ which is directly proportional to exp $(N)$. So the perturbation along this axis grows monotonically with the increase of $N$. When $N$ tends to infinity, the perturbation also grows equally towards infinity. From Figure 8, we notice that the perturbation along the $x$-axis ($\epsilon_x$) exponentially increases as $N$ increases for both the fixed points $R$ and $S$ for all $\gamma$ in [0,2]. So, from the graphical approach, we have $R$ and $S$ to be unstable fixed points. The Jacobian matrices at the fixed points $R$ and $S$ give the eigenvalues $(4, -3\gamma + 6)$ and $\pm\sqrt{((\gamma - 0.72)(\gamma + 7.38))}$, respectively. As the upper limit of $\gamma$ for any fluid is 2, we get both eigenvalues as positive which indicate that the fixed point $R$ behaves as a simple repelling node for $0 \leq \gamma \leq 2$. This is also vividly seen from the phase plot shown in Figure 9. When $\gamma = 2$, one of the eigenvalues of $J_R$ becomes zero and the sign of the remaining eigenvalue is positive. So $R$ is nonhyperbolic and unstable. So from both approaches, $R$ comes out to be unstable for all values of $\gamma$ with $\Omega_\phi = 0$. When $\gamma < 2$, the fixed point $R$ represents a past time attractor with both positive eigenvalues that depicts the inflationary epoch of the evolving Universe. Also the eigenvalues of the Jacobian matrix at $S$ take an opposite sign, and hence, $S$ is a saddle and unstable as seen from the perturbation plot as well as the analytical investigations. This represents the matter-dominated phase of the Universe with $\Omega_\phi = 1$.

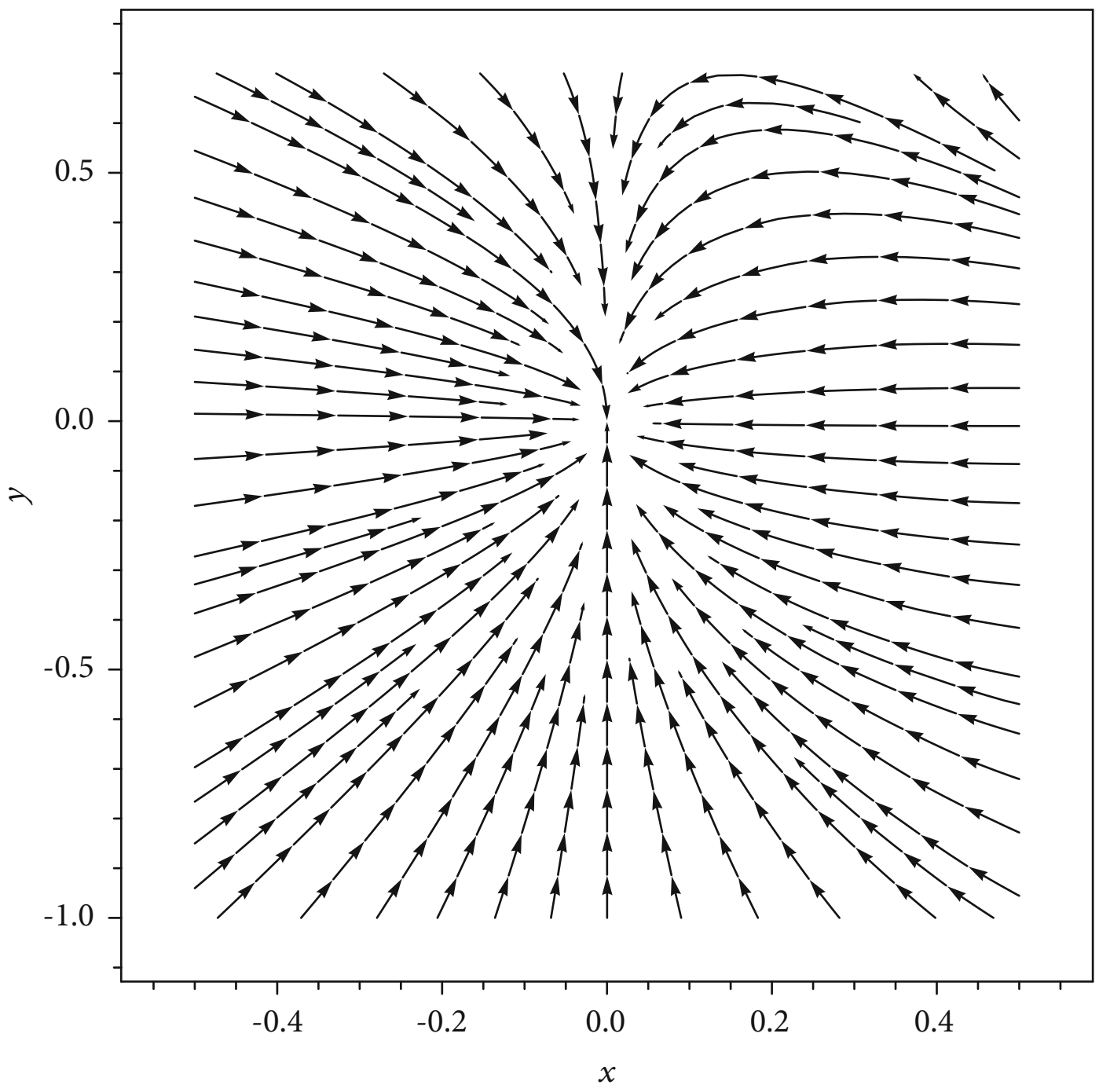

FIGURE 3: Shows the phase plot of the system without potential for $\gamma = 4/3$ for a stable fixed point $P$.

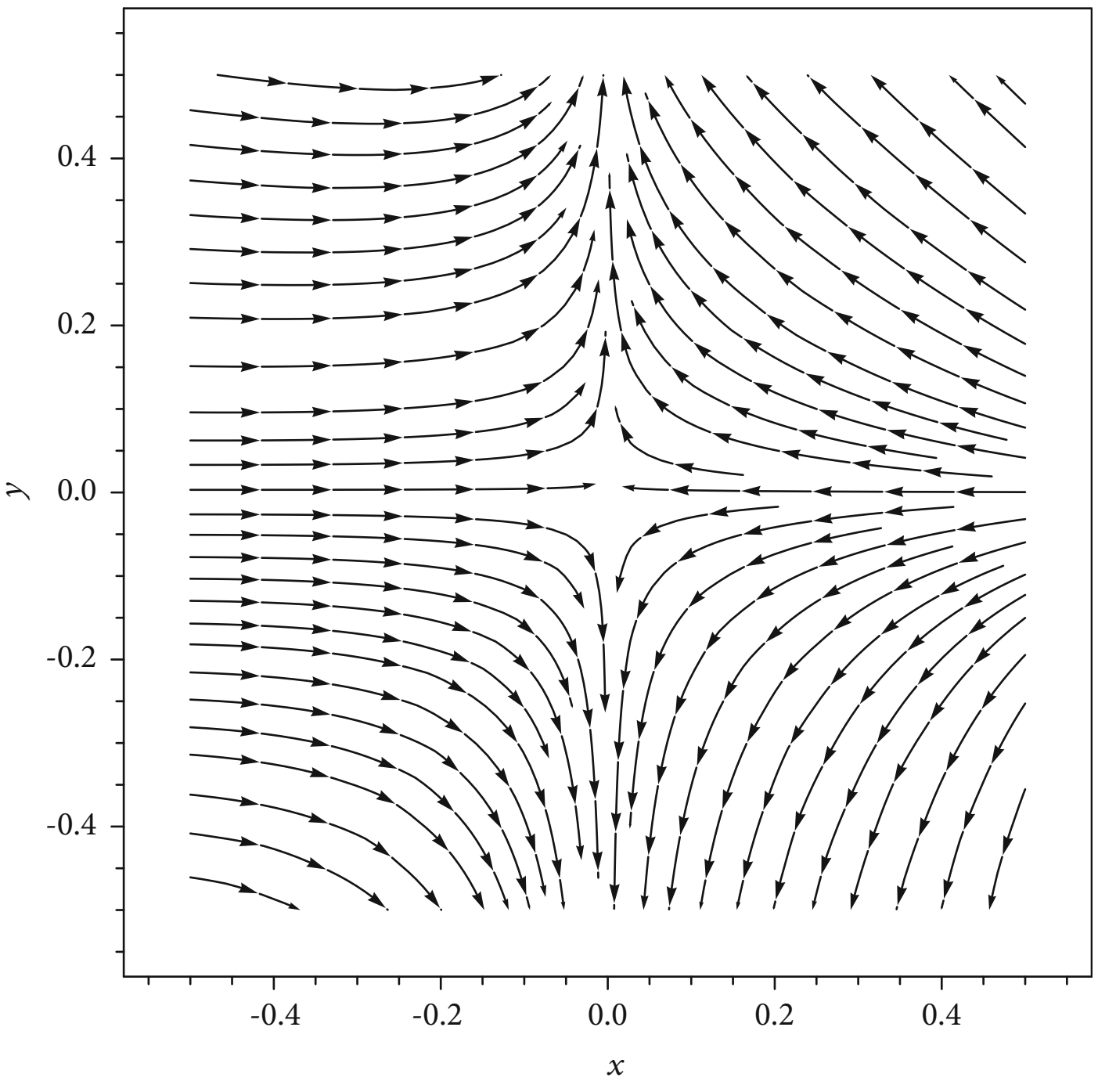

FIGURE 4: Shows the phase plot of the system without potential for $\gamma = 0$ showing that $P\,(x = 0, y \to 0)$ is a saddle fixed point.

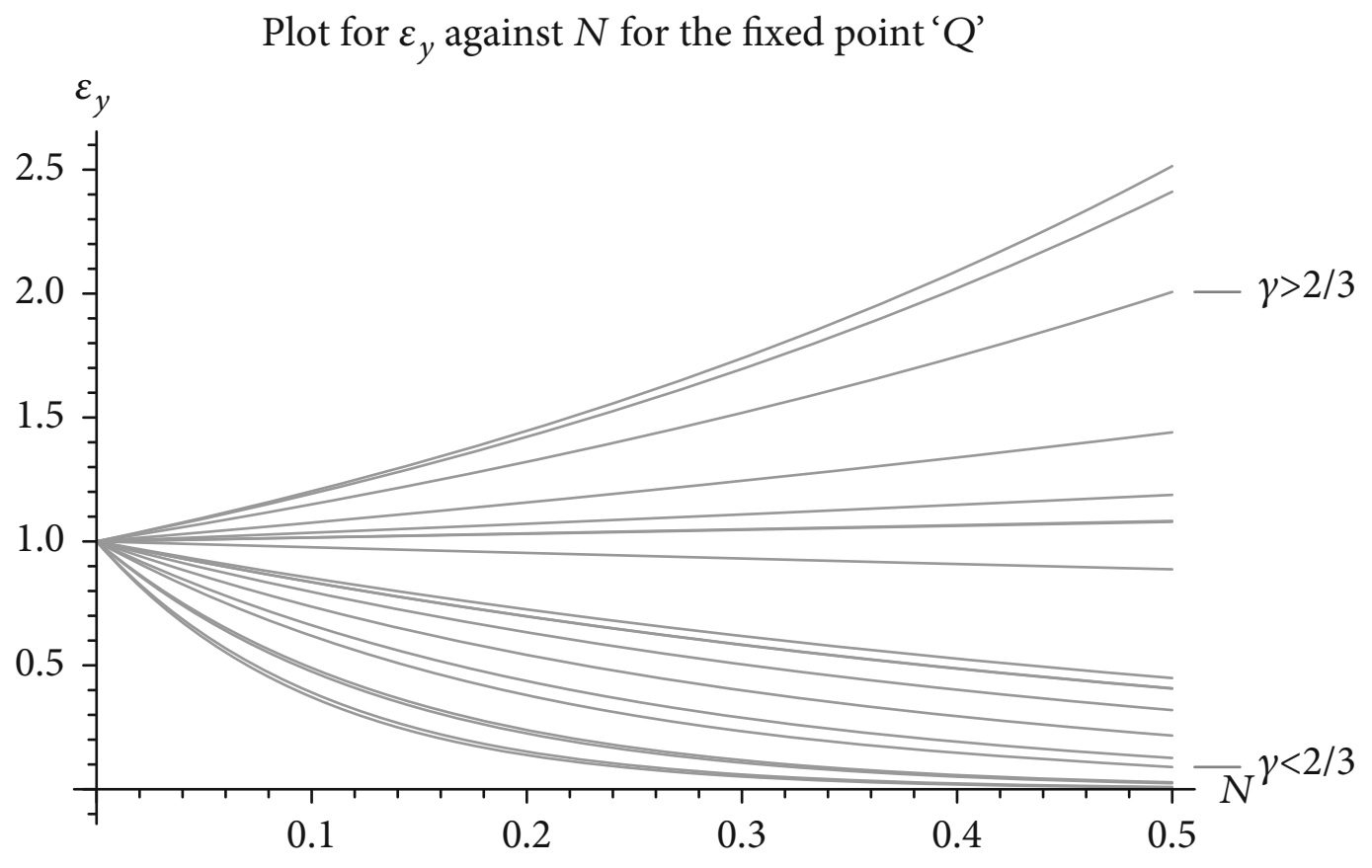

Figure 5: Shows the variation of $\epsilon_y$ against $N$ showing that $Q\,(x = 0, y \to 1)$ is a stable fixed point for $\gamma < 2/3$.

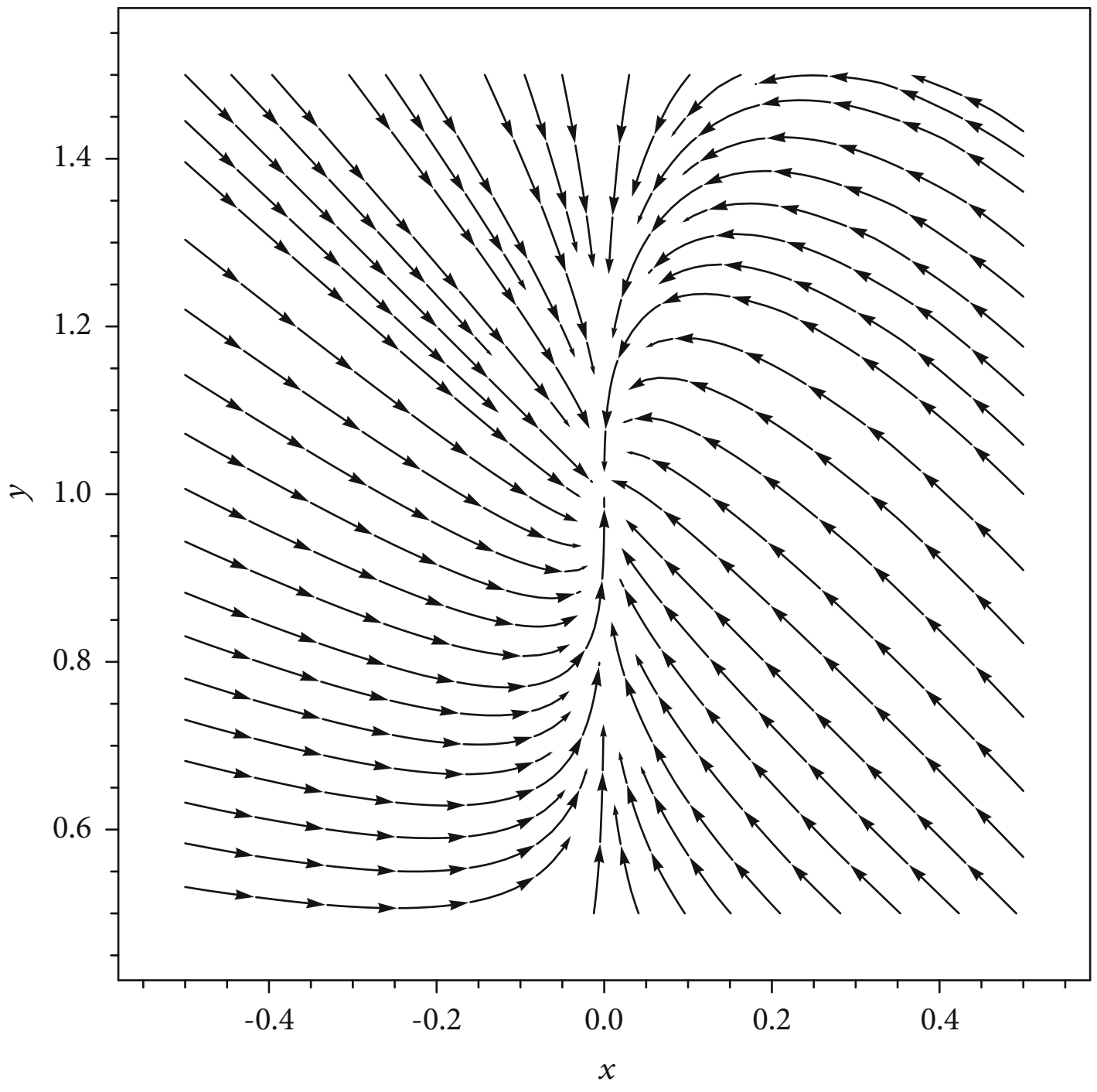

Figure 6: Shows the phase plot of the system without potential for $\gamma = 0$ showing that $Q$ is a stable fixed point.

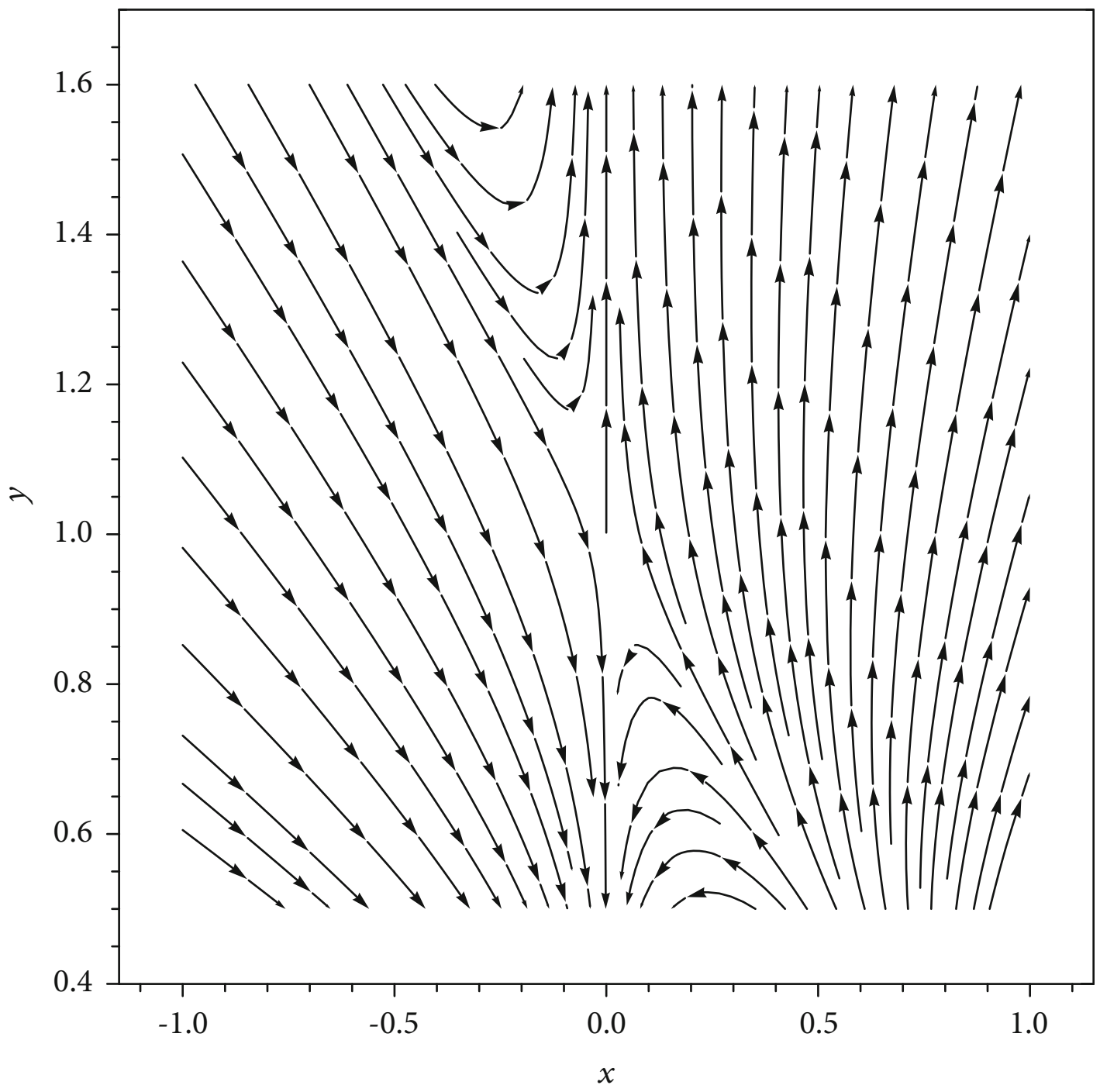

FIGURE 7: Shows the phase plot of the system without potential for $\gamma = 4/3$ showing unstable $Q\,(x = 0, y \to 1)$ when $\gamma > 2/3$.

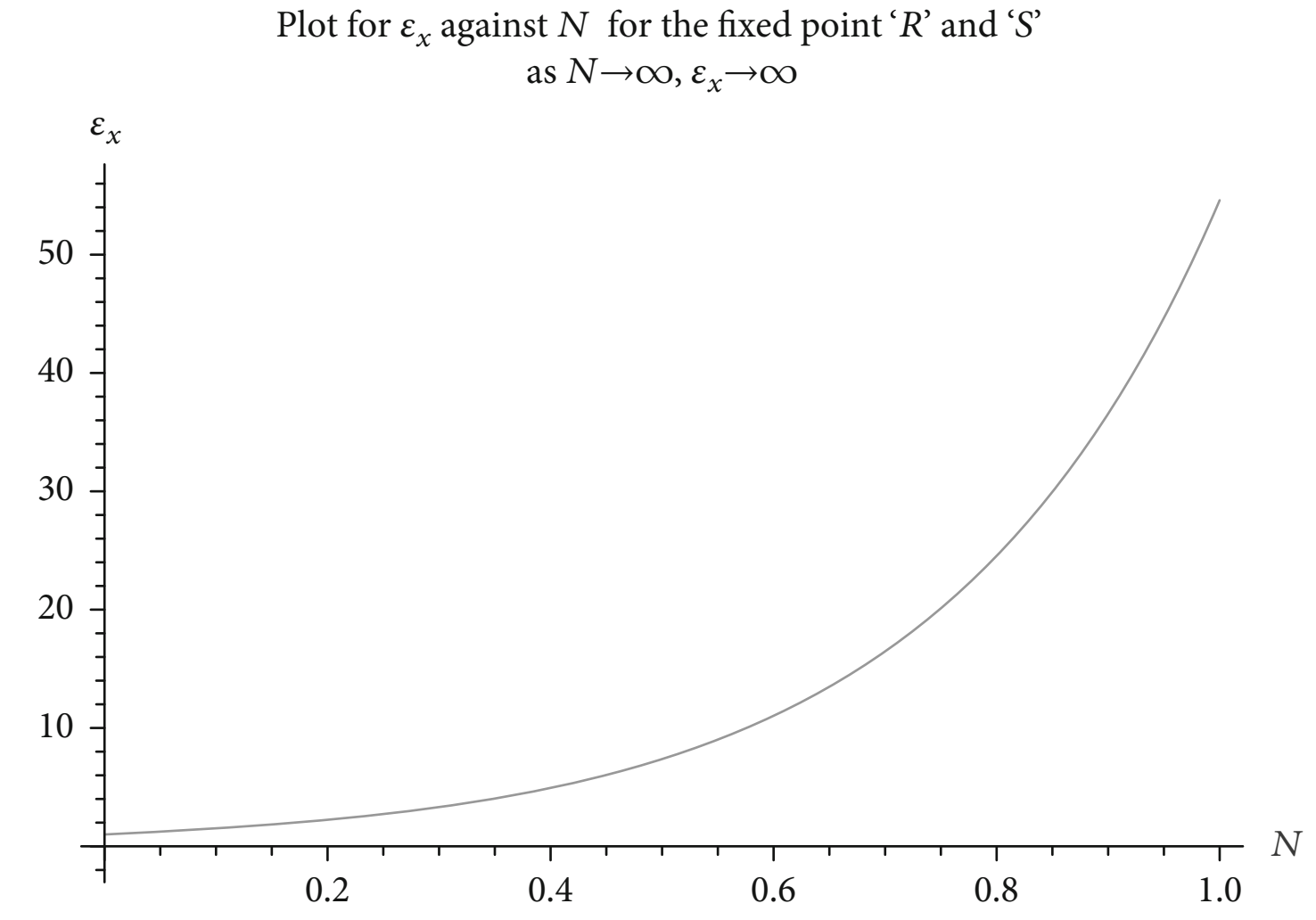

FIGURE 8: Shows the variation of perturbations along the $x$-axis ($\epsilon_x$) for fixed points $R$ and $S$.

*2.2. Analysis with Potential.* The study of a scalar field coupled with potential energy in a dynamical system approach has many applications in General Relativity, specially to explain several cosmological features. We take different potential forms to analyze the system.

*2.2.1. Analysis with Simple Exponential Potential.* We first take simple exponential potential of the form $V(\phi) = V_o e^{-\lambda\phi}$ where $V_o > 0$ is a constant and $\lambda$ is a real constant. This is a common functional form for the self-interaction potential [43]. We consider the following dimensionless variables as $x^2 = \dot{\phi}^2/6H^2$, $y = \rho/3H^2$, and $z = V/3H^2$.

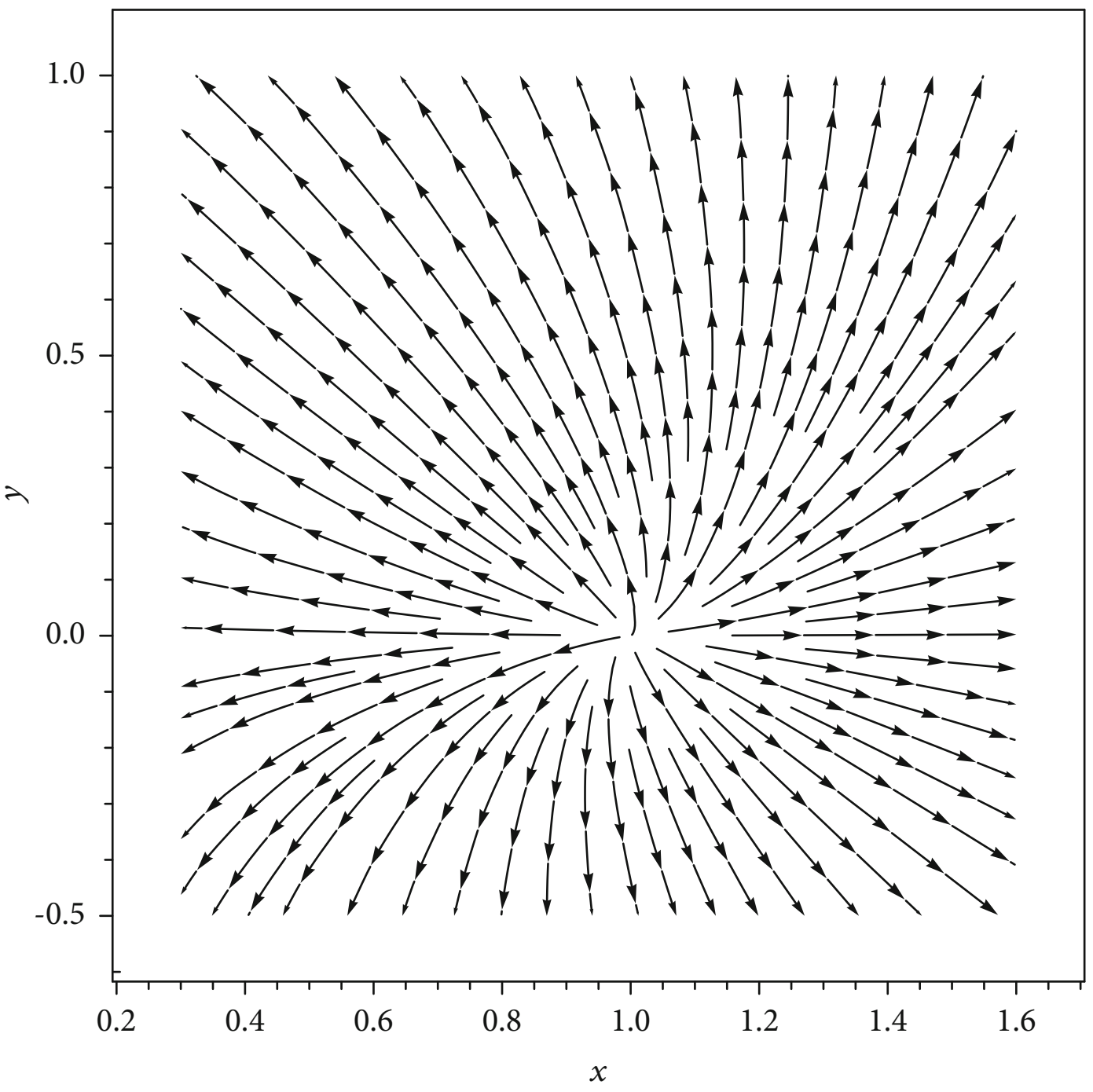

Figure 9: Shows the phase plot for the fixed point $R$.

Table 2: Table for dynamical system with simple exponential potential.

| Fixed points | $x$ | $y$ | $z$ | $\Omega_\phi$ | Stability | $\omega_\phi$ |
|---|---|---|---|---|---|---|
| $A$ | $\sqrt{\frac{2}{3}}\frac{1}{\lambda}$ | 0 | $1-\frac{\lambda^2}{6}$ | $1-\sqrt{\frac{2}{3}}\frac{1}{\lambda}-\frac{\lambda^2}{6}$ | Stable for all $\sqrt{2}<\lambda<\sqrt{3}$ and $\gamma>2/3$ | 0 |
| $B$ | $\frac{\lambda}{\sqrt{6}}$ | 0 | $\frac{4}{3\lambda^2}$ | $\frac{4}{3\lambda^2}+\frac{\lambda}{\sqrt{6}}$ | Stable for all $\lambda<\sqrt{2}$ and $\gamma>2/3$ | -1 |
| $C$ | 1 | 0 | 0 | 1 | Unstable for any $\lambda$ and $\gamma$ | 1 |

With these new variables, the system of equations (1), (6), and (7) reduces to the following set of autonomous system of ordinary differential equations:

$$x' = 2x^3 + \frac{\gamma}{2}xy - xz - 2x + \frac{\sqrt{3}}{2}\lambda z, \quad (14)$$

$$y' = \gamma y^2 + 4yz^2 - 2yz + (2-3\gamma)y, \quad (15)$$

$$z' = -2z^2 - \lambda\sqrt{6}zx + 4zx^2 + \gamma zy + 2z, \quad (16)$$

where the overhead dash represents differentiation w.r.t. logarithmic time ($N = \ln a$) and curvature constant, $k = H^2a^2(x^2+y+z-1)$. The total energy density of the homogeneous scalar field ($\Omega_\phi$) in terms of the newly introduced variables is given by

$$\Omega_\phi = \frac{\left(\dot{\phi}^2/2\right) + V(\phi)}{3H^2} = x + z. \quad (17)$$

Also, the effective equation of state ($\omega_\phi$) for the scalar field for any fluid is given by

$$1 + \omega_\phi = \frac{p_\phi + \rho_\phi}{\rho_\phi} = \frac{2x}{x+z}, \quad (18)$$

where $p_\phi = (1/2)\phi^2 - V$ is the pressure and $\rho_\phi = (1/2)\phi^2 + V$ is the energy density of the scalar field. To determine the fixed points, we equate $x'$, $y'$, and $z'$ to 0 and the fixed points are studied. We tabulate the fixed points and their stability analysis in Table 2. We obtain three fixed points, $A$, $B$, and $C$, taking $\gamma \neq 2/3$ for the fixed points to be in the finite phase plane. In a three-dimensional problem, for nonhyperbolic fixed points, we cannot use the usual linear stability analysis. There are some other famous methods like Center Manifold Theory [23–25], Lyapunov's functions [44, 45], and phase plot or numerical perturbation of solutions about the critical points [46]. However, for nonhyperbolic fixed points, we will use another strategy to find the stability [43]. We will perturb the system by a small amount ($\epsilon$) and study the evolution of perturbations. If the perturbation grows monotonically with the increase of $N$, then the system moves away from the fixed point, and hence, the system is unstable. Otherwise, if the perturbation decays as $N$ tends to infinity, i.e., if the system comes back to the fixed point following the perturbation or if the perturbation evolves to a constant value, then the system is stable. We will study the phase plot of the system near the fixed point. Since, it is difficult to derive conclusions from the 3D phase plot, perturbations on different axes are considered separately by finding out the perturbation along each axis as a function of $N$. We put $x = x_F + \epsilon_x$ in the dynamical system to find perturbation along the $x$-axis as a function of $N$ where $x_F$ denotes the value of $x$ at the fixed point and $\epsilon_x$ denotes the small perturbation along the $x$-axis. With the same procedure, we obtain the perturbations along each axis for the fixed points $A$, $B$, and $C$ as follows:

(1) For the fixed point $A$, the perturbation along each axes are as follows:

$$\varepsilon_x = E_1 \exp\left(-\left(2 - \frac{4}{\lambda^2}\right)N\right),$$
$$\varepsilon_y = E_2 \exp\left((2 - 3\gamma)N\right),$$
$$\varepsilon_z = E_3 \exp\left(-2\left(1 - \frac{\gamma^2}{3}\right)N\right) + \frac{6\left(\lambda^2 - \lambda^4\right)}{36 - 12\lambda^2} \quad (19)$$

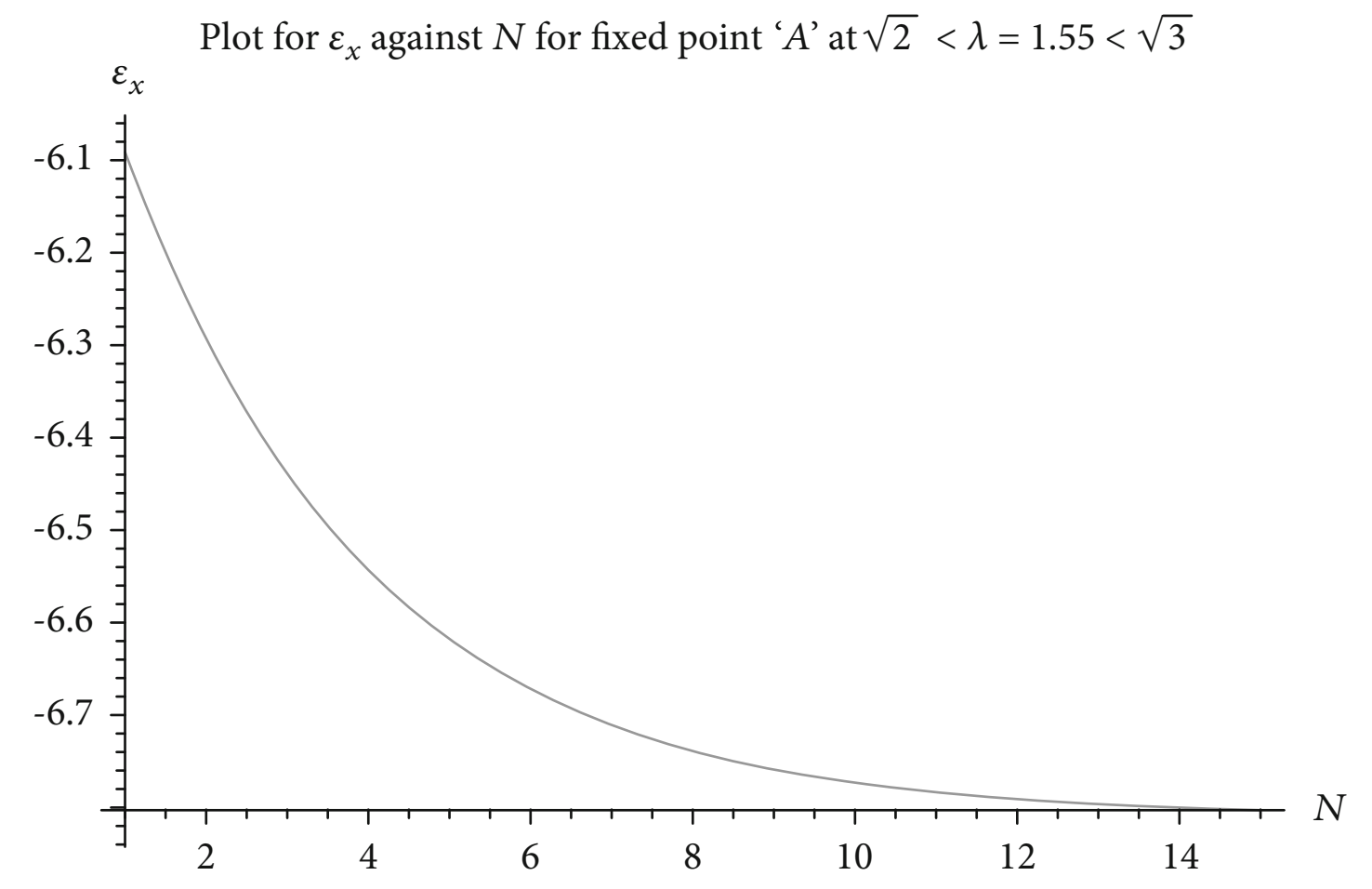

FIGURE 10: Shows the variation of perturbations along the $x$-axis ($\epsilon_x$) against $N$ for fixed point $A$ at $\lambda = 1.55$.

(2) For the fixed point $B$, the perturbation along each axes are as follows:

$$\varepsilon_x = E_4 \exp\left(-\left(2 - \lambda^2\right)N\right) + \frac{\left(\lambda^3 - 6\lambda\right)}{3\sqrt{6\left(2 - \lambda^2\right)}},$$
$$\varepsilon_y = E_5 \exp\left((2 - 3\gamma)N\right),$$
$$\varepsilon_z = E_6 \exp\left(-\left(-2 + \frac{16}{3\lambda^2}\right)N\right) + \frac{8\left(3\lambda^2 - 4\right)}{3\lambda^2\left(16 - 6\lambda^2\right)} \quad (20)$$

(3) For the fixed point $C$, the perturbation function along the $z$-axis is

$$\epsilon_z = E_7 \exp\left(2N\right), \quad (21)$$

where $E_i (1 \le i \le 7)$, $i$ which is taken from the index set $\Xi$, are the arbitrary constants of integration

The perturbation along the $z$-axis is a monotonically increasing function of $N$. So as $N$ increases and tends to infinity, the perturbation long the $z$-axis grows infinitely. Thus, the fixed point $C$ is unstable, and we do not need to study the perturbation along other axes as already the perturbation along the $z$-axis fails to decay or evolve to a constant value as logarithmic time ($N$) increases to infinity. Now, we will plot the perturbation along each axis for each fixed point against logarithmic time ($N$) to graphically determine the nature of stability and its contribution to the behavior of the system.

For the fixed point $A$, the perturbation along the $x$-axis ($\epsilon_x$) comes out to be an exponentially decreasing function of $N$ when the value of $\lambda$ becomes just greater than $\sqrt{2}$. So for those values of $\lambda > \sqrt{2}$, the perturbation along the $x$-axis gradually decreases and finally evolves to a constant value when $N$ tends to infinity. Figure 10 shows how the perturbation along the $x$-axis varies with the increase of $N$ when $\lambda$ lies between $\sqrt{2}$ and $\sqrt{3}$. The perturbation along the $y$-axis ($\epsilon_y$) is an exponential decreasing function of $N$ for $\gamma > 2/3$. So for all $\gamma > 2/3$, the perturbation along the $y$-axis decays to zero as $N$ tends to infinity. Figure 11 shows the variation of $\epsilon_y$ vs. $N$ and Figure 12 shows the variation of perturbation along the $z$-axis ($\epsilon_z$) against $N$. The perturbation function along the $z$-axis ($\epsilon_z$) is an exponentially decreasing function of $N$ when $\lambda$ becomes less than $\sqrt{3}$. It is also seen from Figure 12 that the perturbation along the $z$-axis ($\epsilon_z$) decays to zero as $N$ tends to infinity when $\lambda < \sqrt{3}$. So, when $\sqrt{2} < \lambda < \sqrt{3}$ and $2/3 < \gamma \leq 2$, the perturbation along each axis decays to zero as $N$ increases to infinity which indicates that $A$ is a stable fixed point. We also find the Jacobian matrix at the fixed point $A$ for $\sqrt{2} < \lambda < \sqrt{3}$. We calculate the eigenvalues of the Jacobian matrix at the fixed point with $\lambda = 1.69$. The eigenvalues of the Jacobian matrix are all negative, and the perturbation along each axis eventually decays to zero when $N$ tends to infinity. Hence, $A$ is a stable fixed point. The value of $\Omega_\phi$ has been calculated at the fixed point $A$ and we get $\Omega_\phi = \sqrt{2/3}(1/\lambda) + 1 - (\lambda^2/6)$.

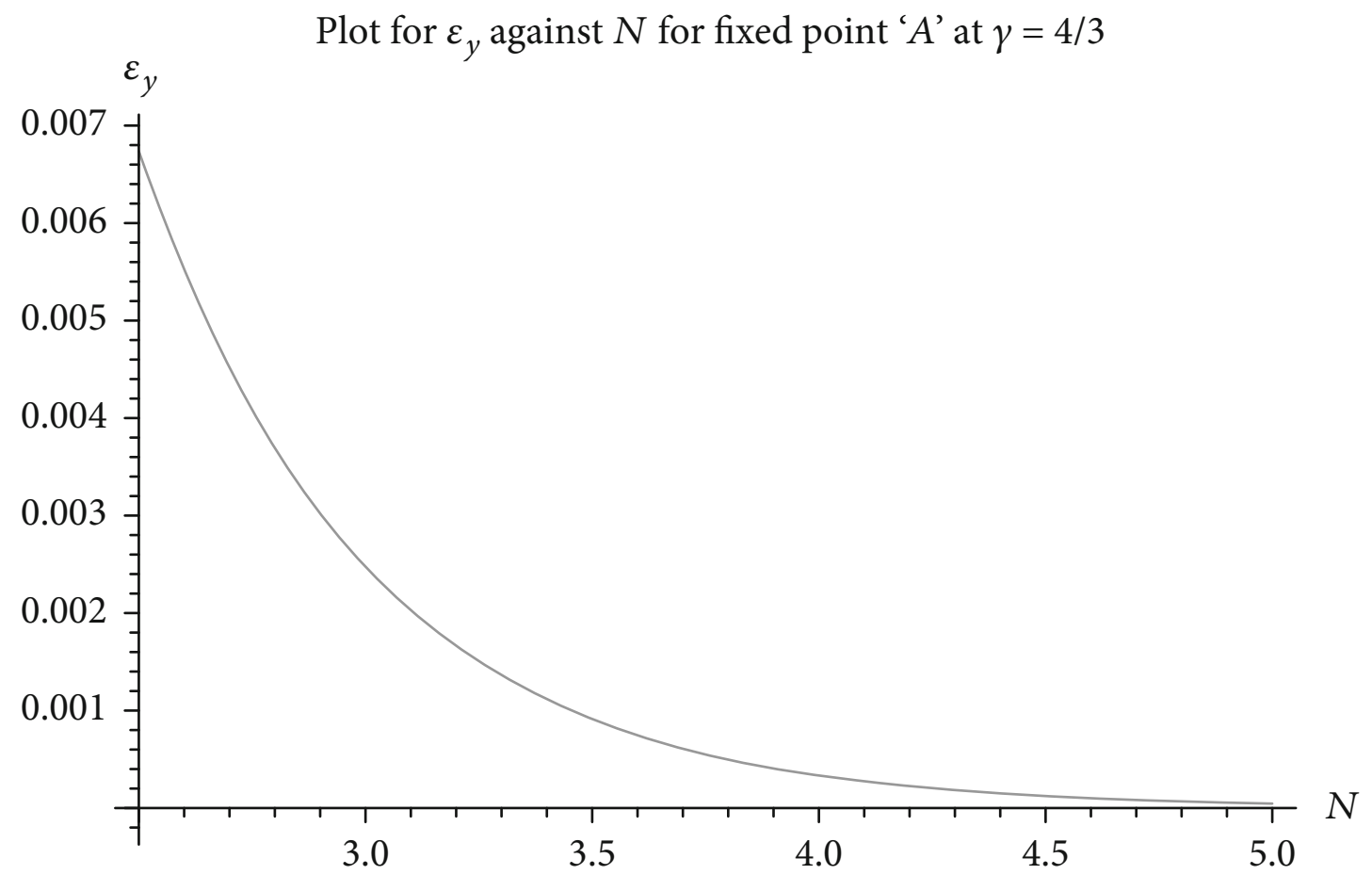

Figure 11: Shows the variation of perturbations along the $y$-axis against $N$ at $\gamma = 4/3$ for the fixed point $A$.

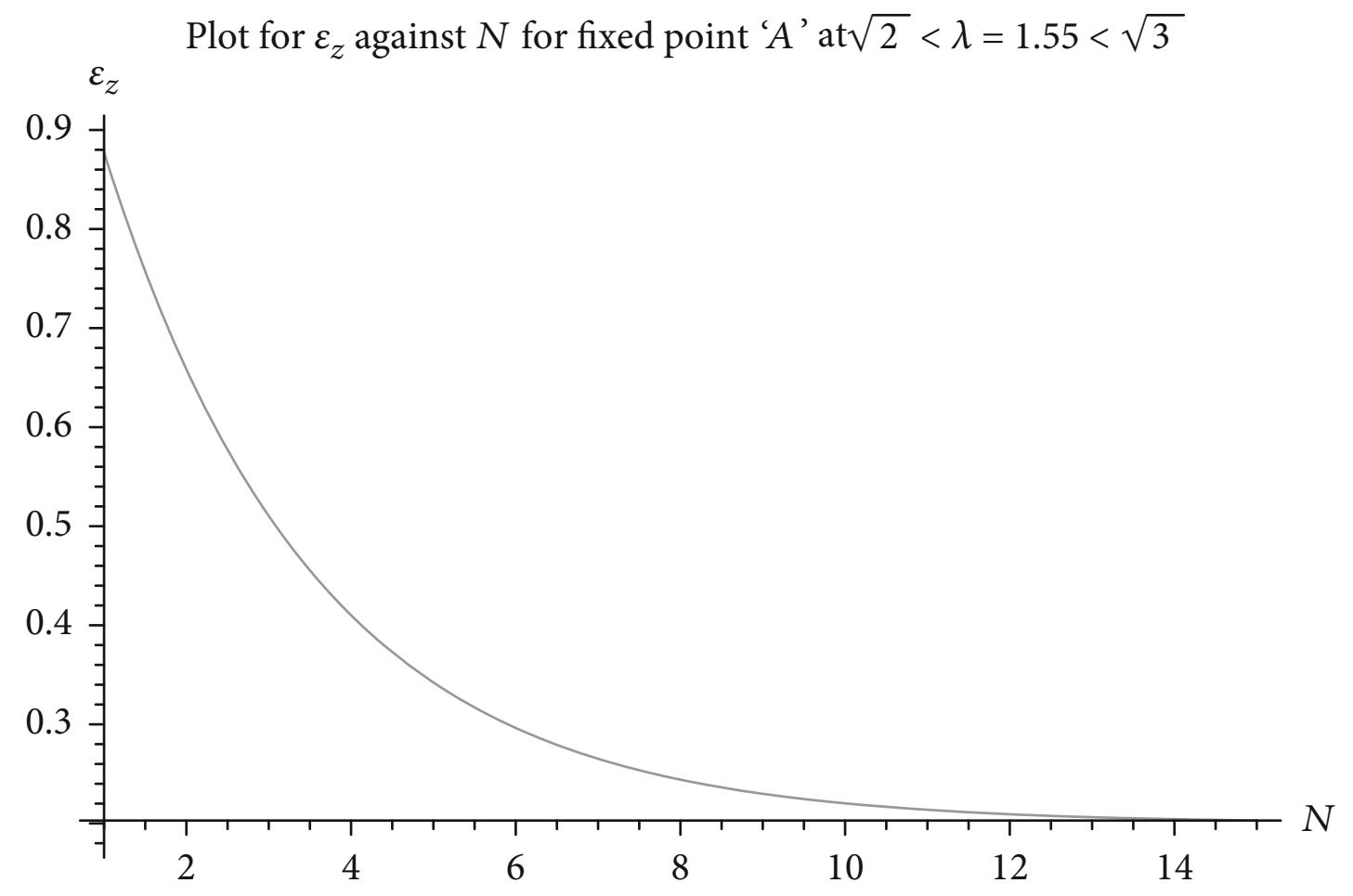

Figure 12: Shows the variation of perturbations along the $z$-axis against $N$ at $\lambda = 1.55$ for the fixed point $A$.

We have evaluated the value of $\Omega_\phi$ at the stable values of $\lambda = 1.69$ determined from the graphical approach. We get the value of $\Omega_\phi$ to be $\Omega_\phi = 1.007$ which agrees with the high precision data given in the Planck 2018 results [46]. From the analytical as well as graphical approach, the fixed point $A$ is a stable fixed point with $\Omega_\phi = 1.007$. Thus, $A$ behaves as a late time attractor with the barotropic fluid index $\gamma$ satisfying $2/3 < \gamma \leq 2$. For fixed point $B$, the perturbation along the $x$-axis ($\epsilon_x$), perturbation along the $y$-axis ($\epsilon_y$), and $\epsilon_z$ decay exponentially to zero as $N$ tends to infinity for $\lambda < \sqrt{2}$ and $\gamma > 2/3$. We have also shown in Figures 13–15, a discussion of the nature of the perturbation functions at each axis. The value of $\Omega_\phi$ at the fixed point $B$ for $\lambda = 0.5 < \sqrt{2}$ has been calculated, and we have found its value as $\Omega_\phi \sim 1$ which is the gross value of $\Omega_\phi$ which as a whole is always 1 [47]. We have observed that the Jacobian matrix at the fixed point $B$ depends on $\lambda$ and evaluated the eigenvalues by finding its characteristic polynomial. Since Figure 15 explains that $B$ is stable for $\lambda < \sqrt{2}$, we find out the Jacobian matrix with $\lambda = 0.5 < \sqrt{2}$ and solve the characteristic polynomial equation. We have found all the eigenvalues to be negative. This indicates that $B$ is a stable fixed point. For the fixed point $C$, the perturbation function along the $z$-axis emerges as an exponentially increasing function of $N$. So with the increase of $N$, the perturbation along the $z$-axis grows exponentially, and henceforth, $C$ is an unstable fixed point. The perturbation plot along the $z$-axis in Figure 16 shows how the perturbation along the $z$-axis rises exponentially with the increase of $N$. At this fixed point $C$, $\Omega_\phi$ is found to be $\Omega_\phi = 1$. We have tabulated all the fixed points along with their stability nature in Table 2. From the effective equation of state for the scalar field derived in terms of $x$ and $y$, we have calculated the value of $\omega_\phi$ at the fixed points $A$, $B$, and $C$. It is shown in Table 2. The Jacobian matrix ($J_A$) for the system at the fixed point $A$ for any $\lambda$ and $\gamma$ is as follows:

$$J_A = \begin{bmatrix} -3 + \dfrac{4}{\lambda^2} + \dfrac{\lambda^2}{6} & \dfrac{\gamma}{\sqrt{6}\lambda} & \sqrt{\dfrac{3}{2}}\lambda - \sqrt{\dfrac{2}{3}}\dfrac{1}{\lambda} \\ 0 & \dfrac{8}{3\lambda^2} + \dfrac{\lambda^2}{3} - 3\gamma & 0 \\ \dfrac{\lambda^3}{\sqrt{6}} - \lambda\sqrt{6} + 8\sqrt{\dfrac{2}{3}}\dfrac{1}{\lambda} - \dfrac{4}{3}\sqrt{\dfrac{2}{3}}\dfrac{1}{\lambda} & \gamma\left(1 - \dfrac{\lambda^2}{6}\right) & \dfrac{8}{3\lambda^2} - 4 \end{bmatrix}. \quad (22)$$

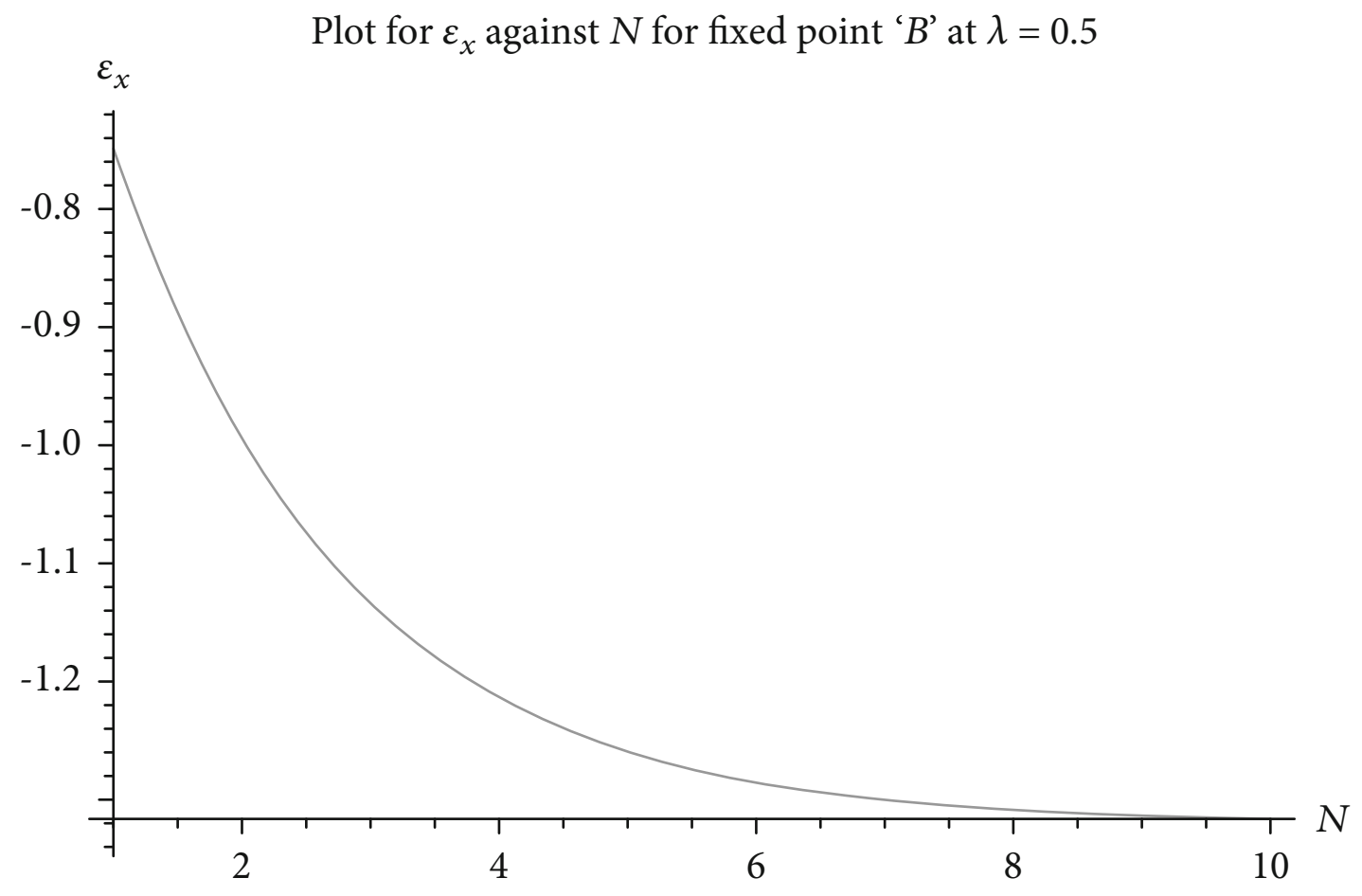

Figure 13: Shows the variation of perturbations along the $x$-axis ($\epsilon_x$) against $N$ at $\lambda = 0.5$ for fixed point $B$.

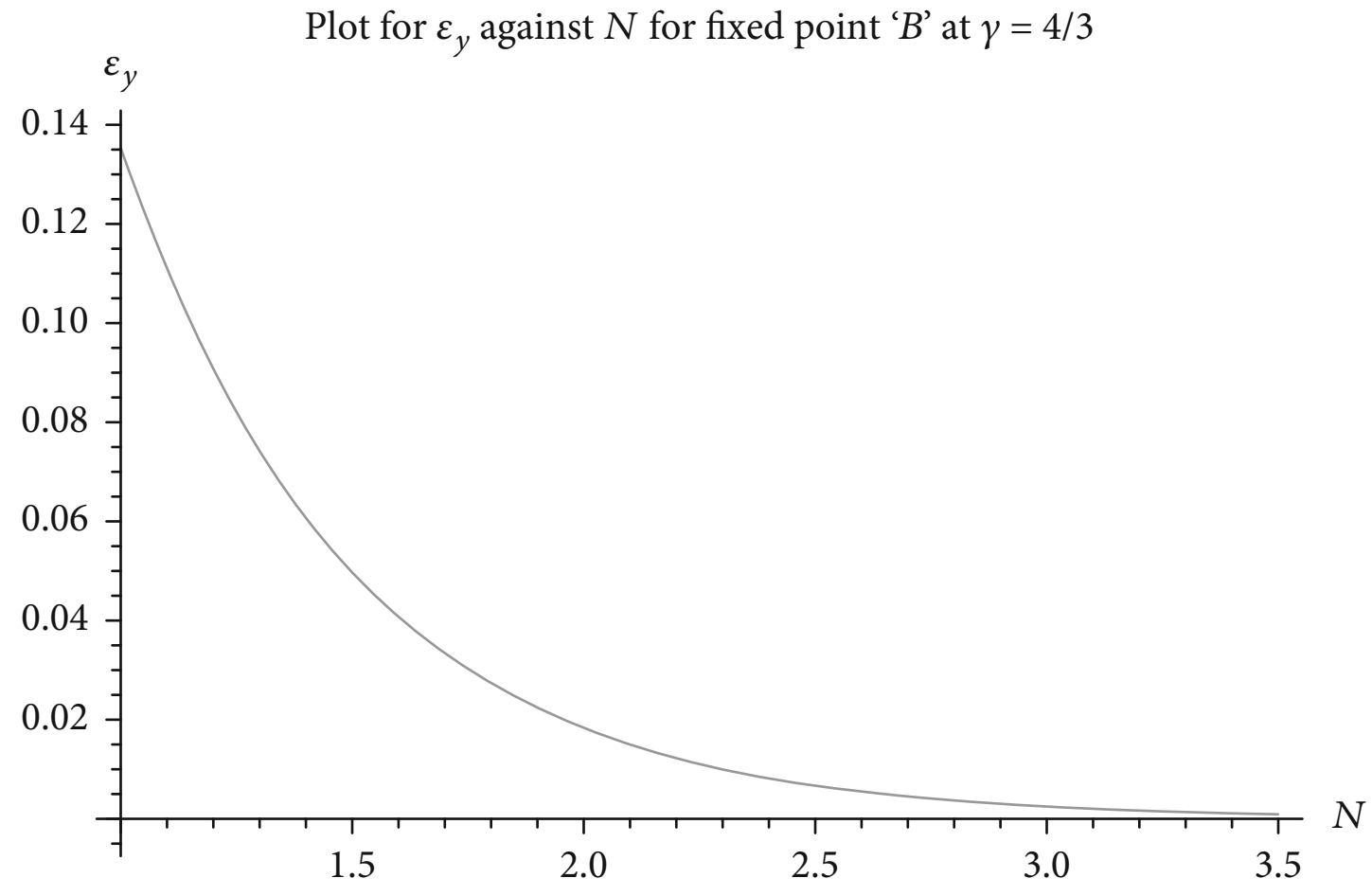

Figure 14: Shows the variation of perturbations along the $y$-axis ($\epsilon_y$) against $N$ at $\gamma = 4/3$ for fixed point $B$.

The Jacobian matrix ($J_B$) at the fixed point $B(0,0,0)$ for any $\lambda$ and $\gamma$ is given by

$$J_B = \begin{bmatrix} \lambda^2 - \dfrac{4}{3\lambda^2} - 2 & \dfrac{\gamma\lambda}{2\sqrt{6}} & \sqrt{\dfrac{3}{2}}\lambda - \dfrac{\lambda}{\sqrt{6}} \\ 0 & \dfrac{2\lambda^2}{3} + (2 - 3\gamma) - \dfrac{8}{3\lambda^2} & 0 \\ \dfrac{32}{3\sqrt{6}}\dfrac{1}{\lambda} & \dfrac{4\gamma}{3\lambda^2} & -2 - \dfrac{\lambda^2}{3} \end{bmatrix}. \quad (23)$$

*2.2.2. Analysis with Double Exponential Potential Form.* In this case, $V(\phi)$ is taken to be $V(\phi) = V_o \exp(-\beta \exp(\sqrt{2}\alpha\phi))$, where $\alpha$ and $\beta$ are positive real constants. Then, we have $dV/d\phi = d/d\phi(V_o \exp(-\beta \exp(\sqrt{2}\alpha\phi))) = -V\sqrt{2}\alpha\beta \exp(\sqrt{2}\alpha\phi) = -V\lambda$ where $\lambda = \sqrt{2}\alpha\beta \exp(\sqrt{2}\alpha\phi)$. It is found that $\lambda$ depends on time. So $\lambda$ can no longer be treated as a constant, but it should be treated as an independent variable. As $\lambda$ becomes an independent variable, the system becomes a four-dimensional system with four independent variables $x$, $y$, $z$, and $\lambda$ [48, 49]. Using the substitutions of dimensionless variables of $x$, $y$, and $z$ as in the case of the simple exponential form along with $\lambda = \sqrt{2}\alpha\beta \exp(\sqrt{2}\alpha\phi)$ in equations (1), (6), and (7), we obtain a system of differential equations given by equations (14)–(16) along with the following equation:

$$\lambda' = \sqrt{12}\alpha x\lambda. \tag{24}$$

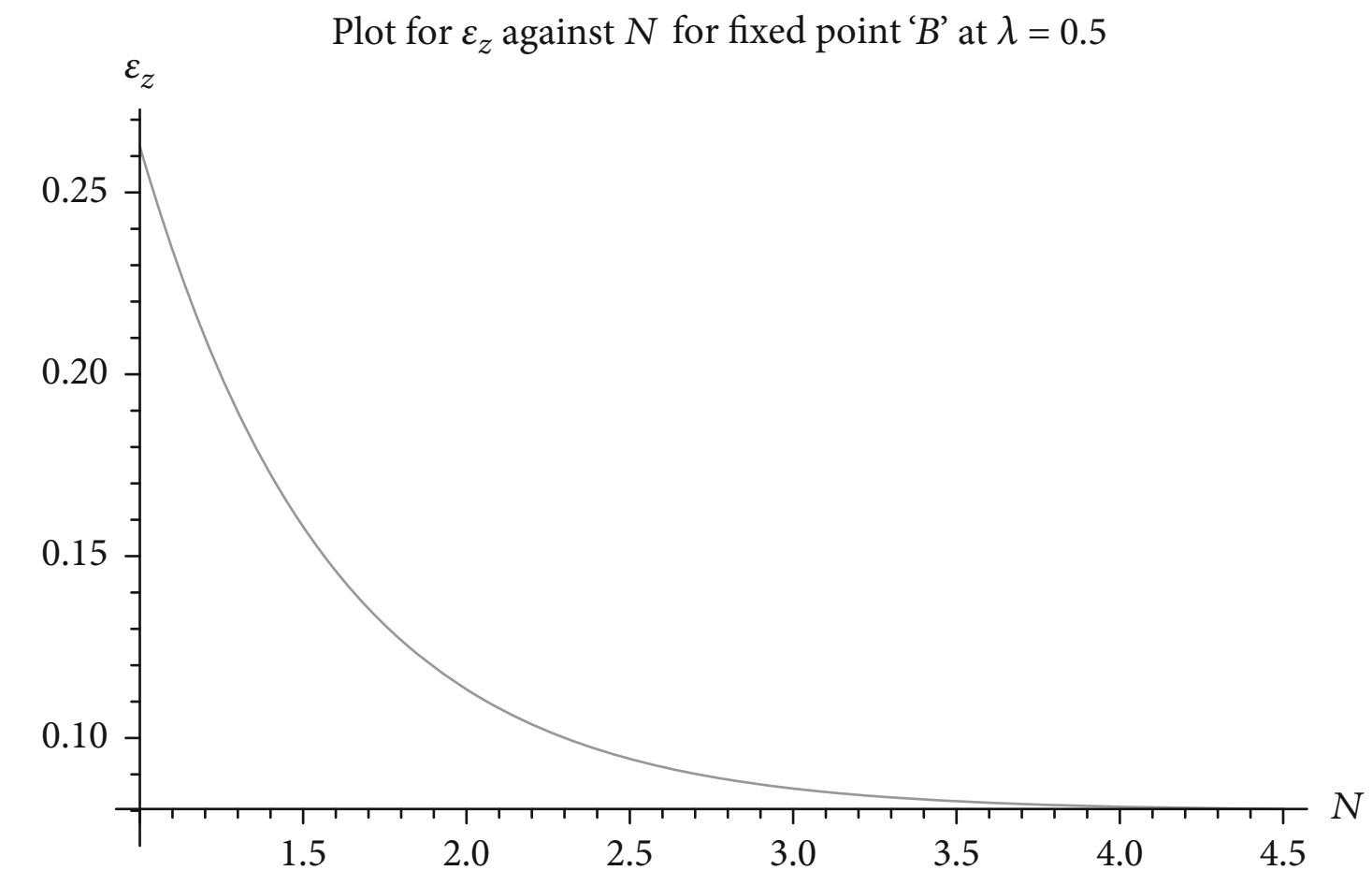

Figure 15: Shows the variation of perturbations along the $z$-axis, $\epsilon_z$ against $N$ at $\lambda = 0.5$ for fixed point $B$.

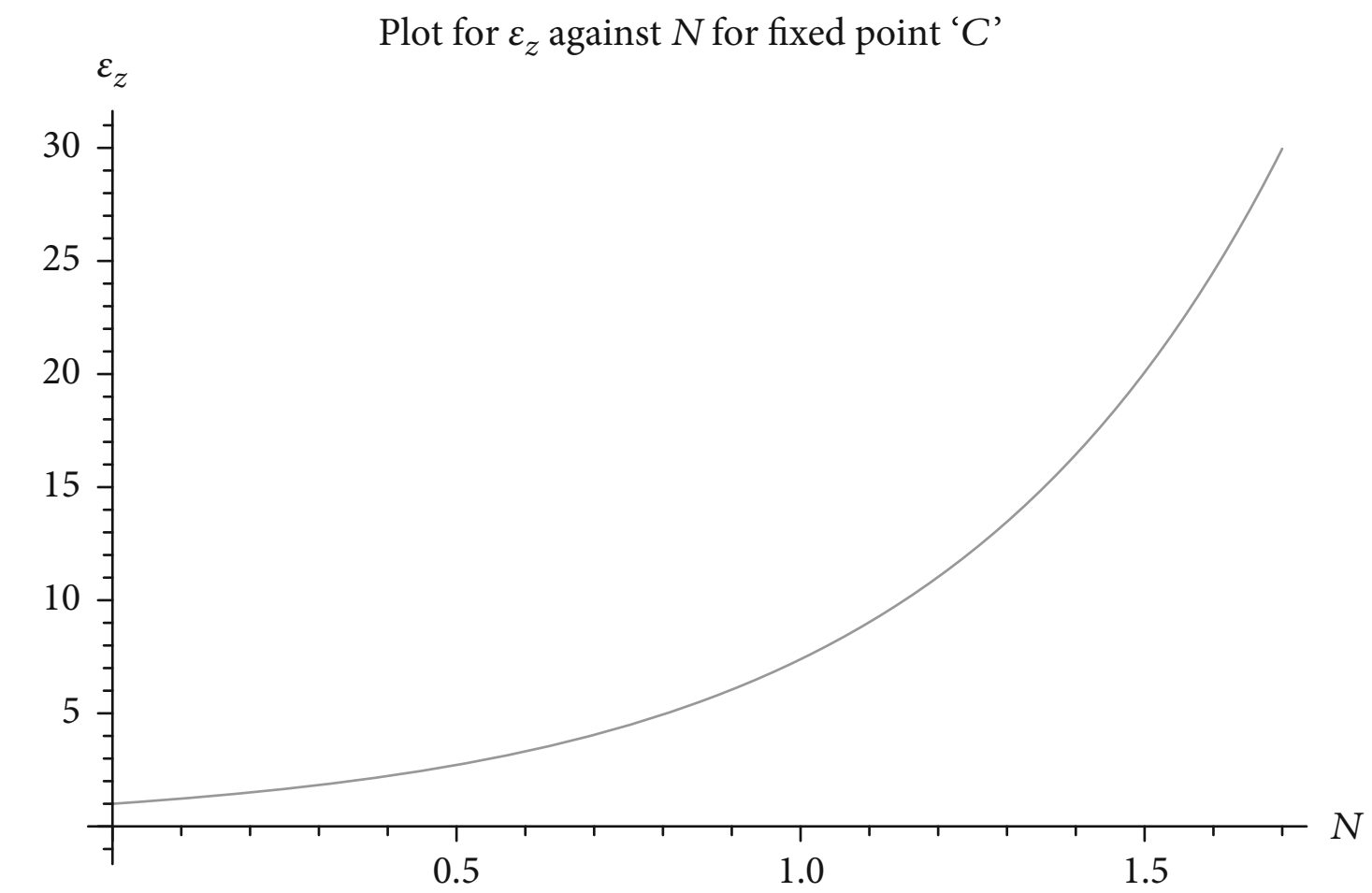

Figure 16: Shows the perturbations along the $z$-axis against $N$ for fixed point $C$.

To find the fixed points, we equate $x'$, $y'$, $z'$, and $\lambda'$ to 0. The fixed points and corresponding cosmological parameters are shown in Table 3. The Jacobian matrices ($J$) for any fixed point $F(x, y, z, \lambda)$ are given by

$$J_F = \begin{bmatrix} 6x^2 + \dfrac{\gamma y}{2} - z - 2 & \dfrac{\gamma x}{2} & \sqrt{\dfrac{3\lambda}{2}} - x & \sqrt{\dfrac{3z}{2}} \\ 8xy & 2\gamma y + 4x^2 - 2z + (2 - 3\gamma) & -2y & 0 \\ \lambda\sqrt{6}z + 8xz & \gamma z & -4z - \lambda\sqrt{6}x + 4x^2 + \gamma y + 2 & -\sqrt{6}zx \\ \sqrt{12}\alpha\lambda & 0 & 0 & \sqrt{12}\alpha x \end{bmatrix}. \tag{25}$$

Table 3: Table for dynamical system with double exponential potential form.

| Fixed points | $x$ | $y$ | $z$ | $\lambda$ | Eigenvalues | $\Omega_\phi$ | Stability | Behavior | $\omega_\phi$ |
|---|---|---|---|---|---|---|---|---|---|
| $F_1$ | 0 | 0 | 1 | 0 | $0, -2, -3, -3\gamma$ | 1 | Stable for all $\gamma \in [0, 2]$ | Stable, late time attractor | -1 (dark energy) |
| $F_2$ | 1 | 0 | 0 | 0 | $4, 4, 6, \sqrt{12}\alpha$ | 1 | Unstable for all $\gamma$ | Unstable | 1 (stiff fluid) |

The Jacobian matrices $J_{F_1}$ and $J_{F_2}$ at the respective fixed points $F_1$ and $F_2$ are given as follows:

$$J_{F1} = \begin{bmatrix} -3 & 0 & 0 & \sqrt{\frac{3}{2}} \\ 0 & -3\gamma & 0 & 0 \\ 0 & \gamma & -2 & 0 \\ 0 & 0 & 0 & 0 \end{bmatrix},$$
$$J_{F2} = \begin{bmatrix} 4 & \frac{\gamma}{2} & -1 & 0 \\ 0 & 4 & 0 & 0 \\ 0 & 0 & 6 & 0 \\ 0 & 0 & 0 & \sqrt{12}\alpha \end{bmatrix}. \tag{26}$$

The Jacobian matrix ($J_{F_2}$) at the fixed point $F_2$ is an upper triangular matrix. So the eigenvalues are its diagonal entries: 4, 4, 6, and $\sqrt{12}\alpha$. All the eigenvalues are nonzero positive quantities since $\alpha$ is also a positive real constant. So the fixed point $F_2$ is a hyperbolic fixed point in which all its eigenvalues are positive. Hence, $F_2$ is unstable. The value of $\Omega_\phi$ at $F_2$ is obtained as 1. The effective equation of state parameter ($\omega_\phi$) is found to be $\omega_\phi = 1$. The fixed point $F_2$ shows a kinetic energy-dominated model with a stiff matter fluid. For the fixed point $F_1$, the characteristic polynomial of $J_{F1}$ is $m(m+2)(m+3)(+3\gamma)$, where $m$ denotes the eigenvalue of $J_{F1}$. Solving the characteristic equation, we obtained four eigenvalues which are $m = 0, -1, -2, -3$, and $-3\gamma$. The fixed point $F_1$ is a nonhyperbolic fixed point as one of the eigenvalues takes a zero value. So for stability analysis, we need to find the perturbation ($\epsilon(N)$) along each axis and study the variation of $\epsilon(N)$ with the increase of logarithmic time ($N$). The following are the perturbations along each axis at the fixed point $F_1$:

$$\begin{aligned} \varepsilon_x &= e^{-N} + \sum_{n=1}^{\infty} \left(e^{-2N}\right)^n + 16, \\ \varepsilon_y &= \begin{cases} \dfrac{3}{e^{3\gamma N} - 1}, & \gamma \neq 0, \\ 0, & \gamma = 0, \end{cases} \\ \varepsilon_z &= \frac{e^{-2N}}{1 - e^{-2N}}, \\ \varepsilon_\lambda &= \delta, \end{aligned} \tag{27}$$

where $\delta$ is a constant. The series $\sum_{n=1}^{\infty} (e^{-2N})^n$ is a geometric series with common ratio $e^{-2N} < 1$. So, this series sums to $e^{2N}/(1 - e^{2N})$, and using this sum, $\epsilon_x$ now becomes $\varepsilon_x = e^{-N} + (e^{2N}/(1 - e^{2N})) + 16$ for all real $N$. We want to study the behavior of $\epsilon_x$ as $N \longrightarrow \infty$ where $N$ can take any real value. We can do a sequential approach for this as it is equivalent to finding the limit of a sequence $<S_N>$ whose $N^{\text{th}}$ term is $S_N = e^{-N} + (e^{2N}/(1 - e^{2N})) + 16$. When we take the limit on $S_N$ as $N \longrightarrow \infty$ where $N \in \mathbf{R}$ the set of real numbers, we obtain the limiting value as 16. So when $N$ tends to infinity, the value of $\epsilon_x$ converges to a constant value which is 16. It means that the perturbation along the $x$-axis ($\epsilon_x$) evolves to a constant value, and as such, $\epsilon_x$ fails to grow as $N \longrightarrow \infty$. From the above expression for $\epsilon_y$ by using the same sequential approach, we see that in both the cases for $\gamma = 0$ or $\gamma \neq 0$, $\epsilon_y$ decays to zero as $N \longrightarrow \infty$. The perturbation along the $z$-axis ($\epsilon_z$) also decays to zero when $N \longrightarrow \infty$. Since the perturbation along the $\lambda$-axis is a constant function, the perturbation never grows along the $z$-axis. So the system is stable. Figures 17–19 show the variation of $\epsilon_x$, $\epsilon_y$, and $\epsilon_z$ against $N$ for $F_1$, respectively. We solve the value of $\Omega_\phi$ and it is obtained as 1 [47]. The value of the effective equation of the state parameter ($\omega_\phi$) is also obtained, and we get $\omega_\phi = -1$. The study of the perturbation along each axis by analytical as well as geometrical approach shows that the fixed point $F_1$ is a stable fixed point which behaves as a late time attractor. This fixed point contributes with a dark energy model ($\omega_\phi = -1$).

*2.2.3. Analysis with Inverse Power Law Potential Form.* We take $V(\phi)$ to be $V(\phi) = V_o \phi^{-\alpha}$, where $\alpha$ is a real constant. Then, $dV/d\phi = -\alpha V_o \phi^{-\alpha-1} = -V(\alpha/\phi) = -V\lambda$, where $\lambda = \alpha/\phi$. Here also, since $\lambda$ becomes an independent variable, we get a four-dimensional system with independent variables as $x$, $y$, $z$, and $\lambda$. Using $\lambda = \alpha/\phi$, the expression for $dV/d\phi$, and the dimensionless variables as in the case of simple exponential potential form, the system of equations (1), (6), and (7) reduces to the system of differential equations as in equations (14)–(16) along with the following equation:

$$\lambda' = \frac{-\sqrt{6} x \lambda^2}{\alpha}. \tag{28}$$

The Jacobian matrix ($J$) at any fixed point $G(x, y, z, \lambda)$ is given by

$$J_G = \begin{bmatrix} 6x^2 + \frac{\gamma y}{2} - z - 2 & \frac{\gamma x}{2} & \sqrt{\frac{3\lambda}{2}} - x & \sqrt{\frac{3z}{2}} \\ 8xy & 2\gamma y + 4x^2 - 2z + (2 - 3\gamma) & -2y & 0 \\ \lambda\sqrt{6}z + 8xz & \gamma z & -4z - \lambda\sqrt{6}x + 4x^2 + \gamma y + 2 & -\sqrt{6}zx \\ \frac{-\sqrt{6}\lambda^2}{2} & 0 & 0 & -\sqrt{6}x\lambda \end{bmatrix}. \quad (29)$$

Here, we get the same fixed points $(G_1, G_2)$, i.e., $(G_1 = F_1, F_2 = G_2)$ as in the above case of double exponential potential form. For the fixed point $G_1$, we get the same Jacobian matrix $(J_{G_1} = J_{F_1})$, and so the analysis for $G_1$ remains the same as what was done in the double exponential form for the fixed point $F_1$. The fixed point $G_2$ becomes nonhyperbolic with the following Jacobian matrix $(J_{G_2})$:

$$J_{G_2} = \begin{bmatrix} 4 & \frac{\gamma}{2} & -1 & 0 \\ 0 & 4 & 0 & 0 \\ 0 & 0 & 6 & 0 \\ 0 & 0 & 0 & 0 \end{bmatrix}. \quad (30)$$

The eigenvalues of $J_{G_2}$ are 4, 4,6, and 0. Since one of the eigenvalues becomes zero, $G_2$ becomes nonhyperbolic. To analyze the stability of this fixed point, we find out the perturbation along each of the four axes as a function of $N$ as follows:

$$\begin{aligned} \epsilon_x &= \frac{2\left(e^{4N} - 1\right) \pm \sqrt{(2 - 2e^{4N})^2 - 4e^{4N}(1 - e^{4N})}}{2(1 - e^{4N})}, \\ \epsilon_y &= \begin{cases} \dfrac{6 - 3\gamma}{e^{-(6-3\gamma)N} - \gamma}, & \gamma \neq 0, \\ \varsigma e^{N}, & \gamma = 0, \end{cases} \\ \epsilon_z &= \frac{e^{2N}}{1 + e^{2N}}, \\ \epsilon_\lambda &= \delta', \end{aligned} \quad (31)$$

where $\delta'$ is a constant. As in the above case of the double exponential potential form, using sequential approach, we get $\epsilon_x$ which tends to a constant value as $N \longrightarrow \infty$. When $\gamma \neq 0$, $\epsilon_y$ tends to a unique limit as $N \longrightarrow \infty$. But when $\gamma = 0$, $\epsilon_y$ tends to infinity as $N \longrightarrow \infty$. By using L'Hospital's rule of finding limit, we see that $\epsilon_z$ attains a constant value which is 1 as $N \longrightarrow \infty$. Also $\epsilon_\lambda$ is independent of $N$ as it is a constant function. So $\epsilon_\lambda$ always attains a constant value $(\delta')$ when $N \longrightarrow \infty$. Hence, the fixed point $G_2$ is stable for $\gamma \neq 0$. Figures 20–22 show the perturbation plot of $G_2$ for different axes. We get the value of $\Omega_\phi = 1$ and the effective equation of state parameter $\omega_\phi = 2$ which represents the ultrarelativistic fluid model. Table 4 shows the fixed points and their stability conditions.

## 3. Conclusion

We have presented, in this work, a dynamical system perspective of scalar field FRW cosmology without potential as well as with potential. In the absence of potential, we found a two-dimensional dynamical system and four critical points $P$, $Q$, $R$, and $S$. When $\gamma > 2/3$, since all the eigenvalues of the diagonal matrix $(J_P)$ at the critical point $P$ take negative values, it confirms the stability of the fixed point $P$ which behaves as an attracting node. This is also shown by the phase plot in Figure 3. When $\gamma > 2/3$, the eigenvalues take opposite signs. This contributes to our model with a saddle fixed point which is unstable with $\Omega_\phi = 0$. The phase plot in Figure 4 shows that $P$ is unstable when $\gamma > 2/3$. We have also added the perturbation plots along each axis and studied how they vary with the rise of the logarithmic time $(N)$ for the fixed point $P$. The phase plots have also supported our analytical findings. With the fixed point $P$, our developed cosmological model becomes stable for the ultrarelativistic fluid model $(\gamma = 2)$, matter-dominated model $(\gamma = 1)$, or radiation cosmology $(\gamma = 4/3)$ which describes the cosmic features [50]. Also, the fixed point $Q$ acts as an attracting node with all the eigenvalues of the Jacobian matrix $J_Q$ being less than zero along with $\Omega_\phi = 1$ for $\gamma < 2/3$. Since our system is stable for $\gamma < 2/3$, it is stable for the dark energy model $(\gamma = 0)$. Also, this scalar field-dominated solution shows a late time attractor that represents the accelerated expansion phase of the Universe for $0 \leq \gamma < 2/3$. The fixed point $R$ behaves as a repelling node and is always unstable for any $0 \leq \gamma \leq 2$ which represents the past time attractor solution since both eigenvalues take a positive value. This fixed point contributes to our model with the inflationary epoch of the evolving Universe. The fixed point $S$ always possesses opposite signs of eigenvalues and gives us a saddle fixed point with $\Omega_\phi = 1$. Hence, it is unstable.

When we analyze in the presence of a simple exponential potential, we can extend the system to a three-dimensional dynamical system. Studying a dynamical system in a flat Universe can be a special case of our present model by putting

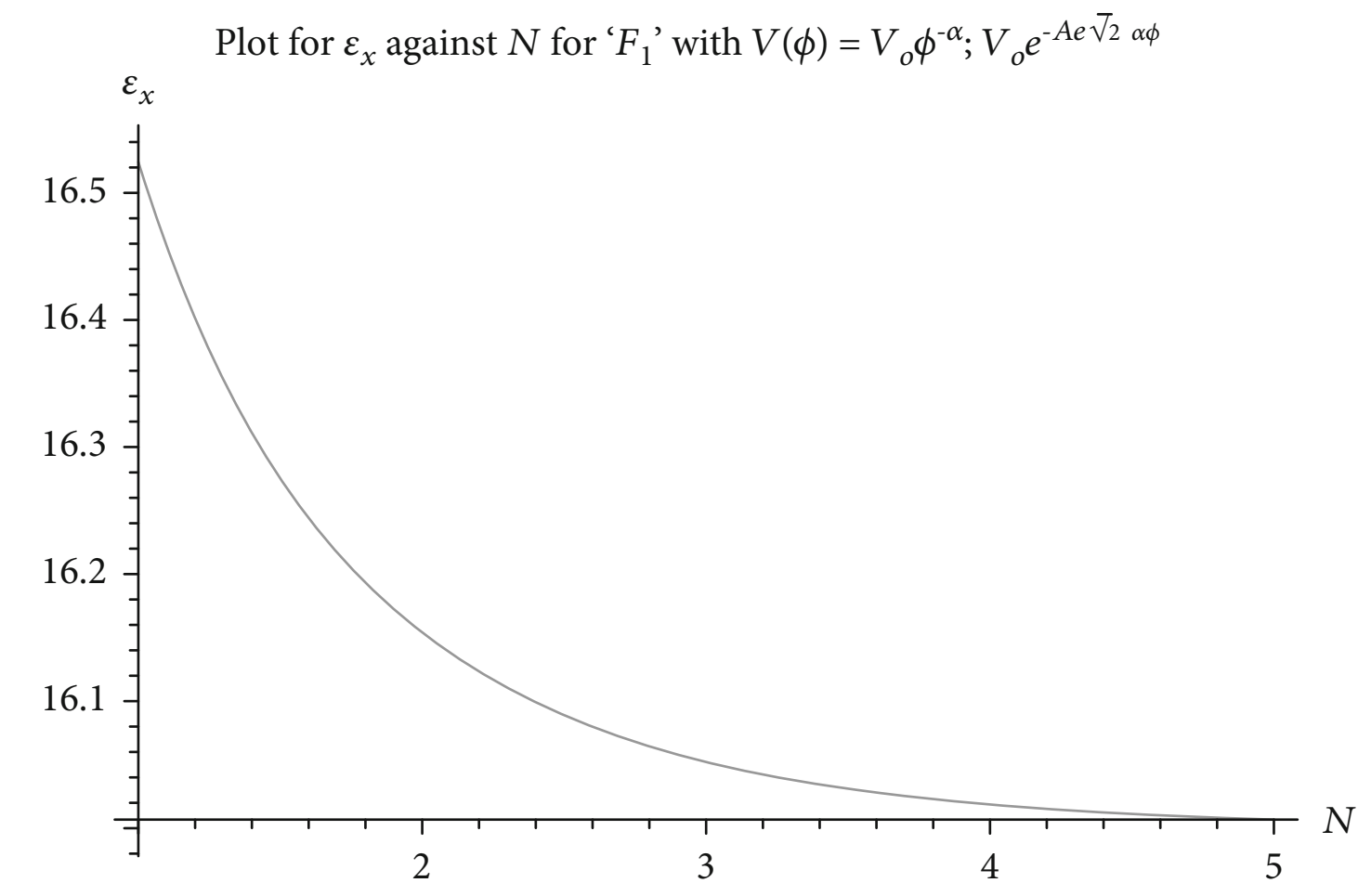

Figure 17: Shows the variation of perturbations along the $x$-axis ($\epsilon_x$) against $N$ at fixed point ($F_1 = G_1$) for exponential potential and inverse power law forms.

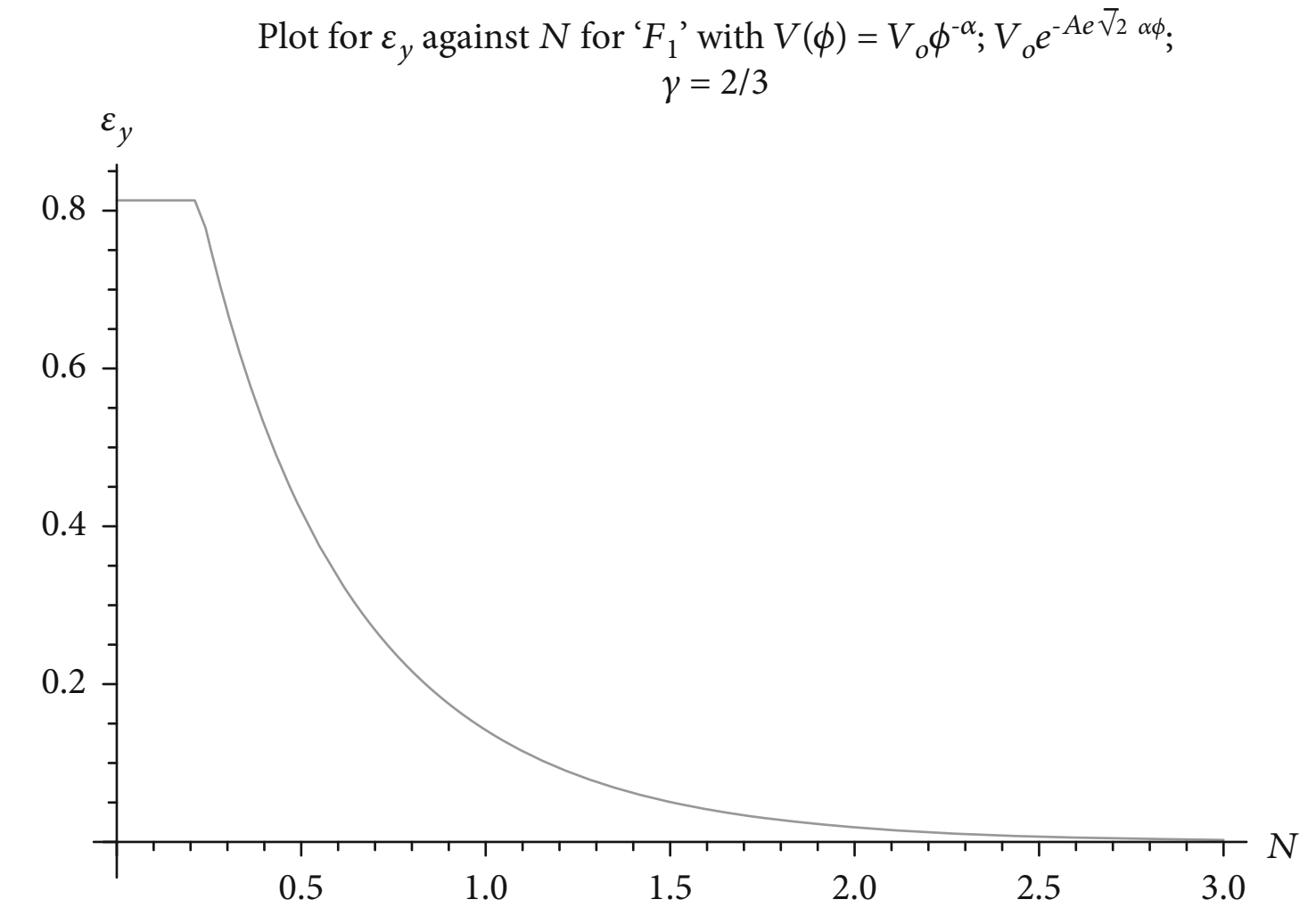

Figure 18: Shows the variation of perturbations along the $y$-axis ($\epsilon_y$) against $N$ at fixed point ($F_1 = G_1$) for exponential potential and inverse power law forms.

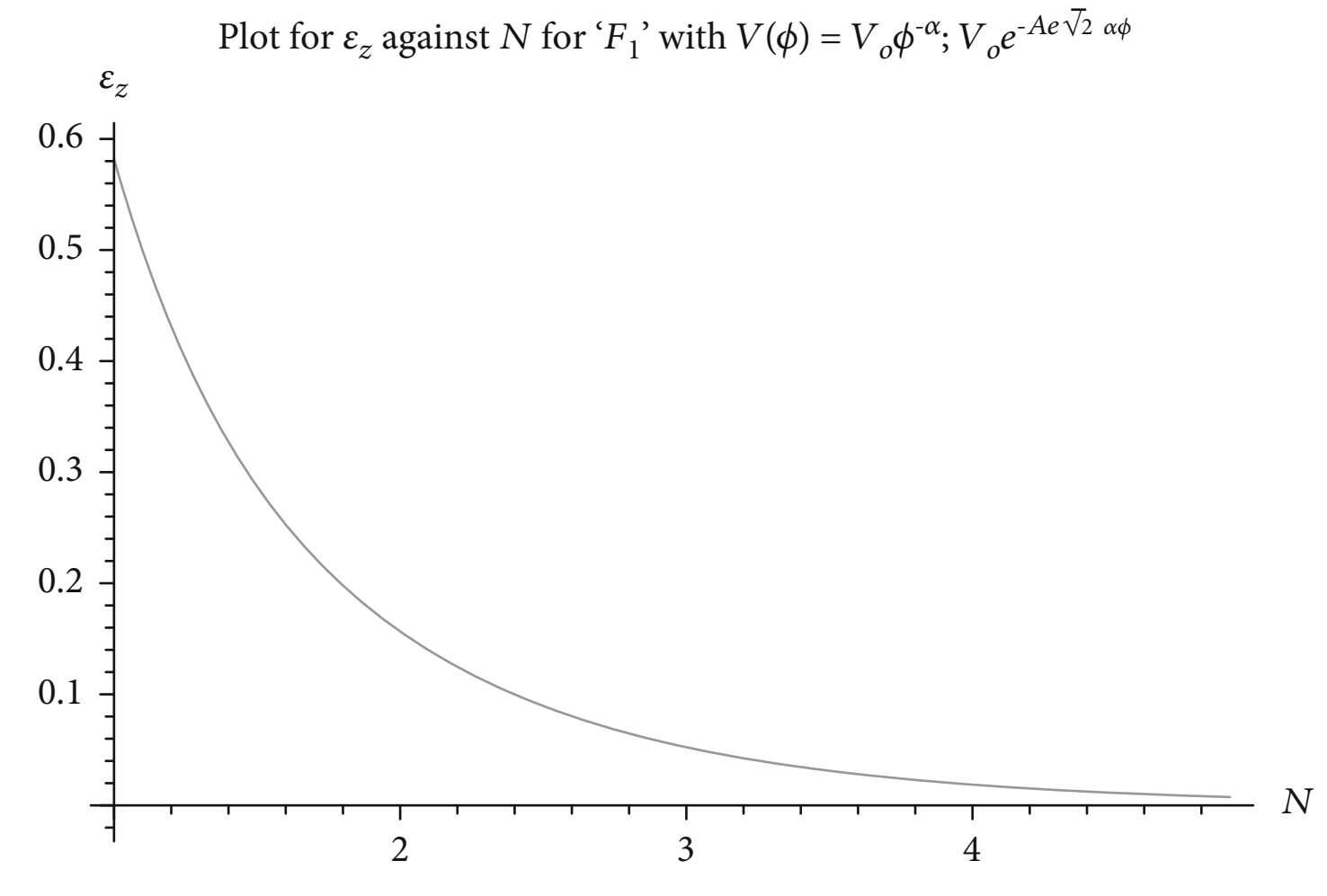

Figure 19: Shows the variation of perturbations along the $z$-axis ($\epsilon_z$) against $N$ at fixed point ($F_1 = G_1$) for exponential potential and inverse power law forms.

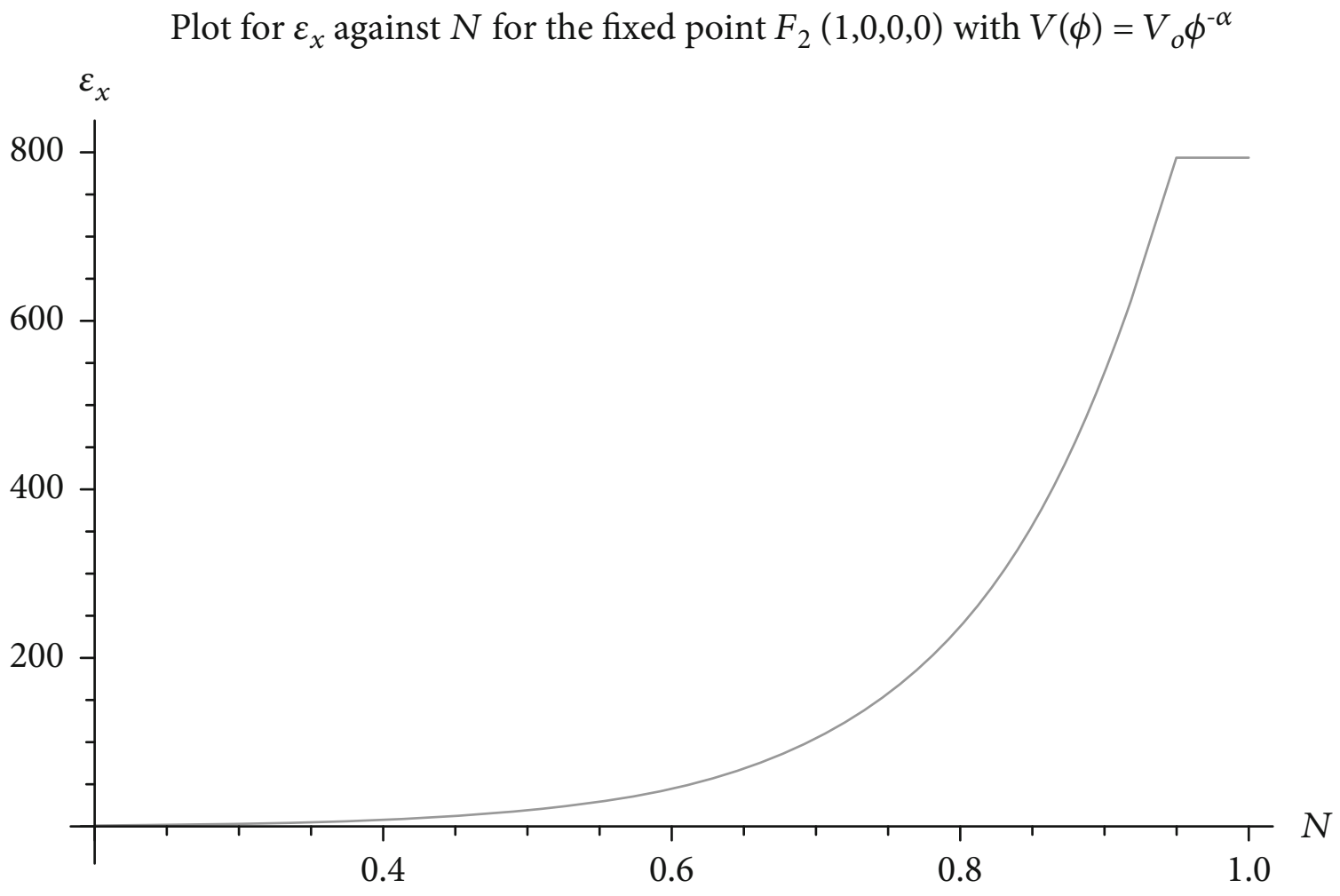

FIGURE 20: Shows the perturbations along the $x$-axis ($\epsilon_x$) against $N$ for $G_2$ using the inverse power law potential form.

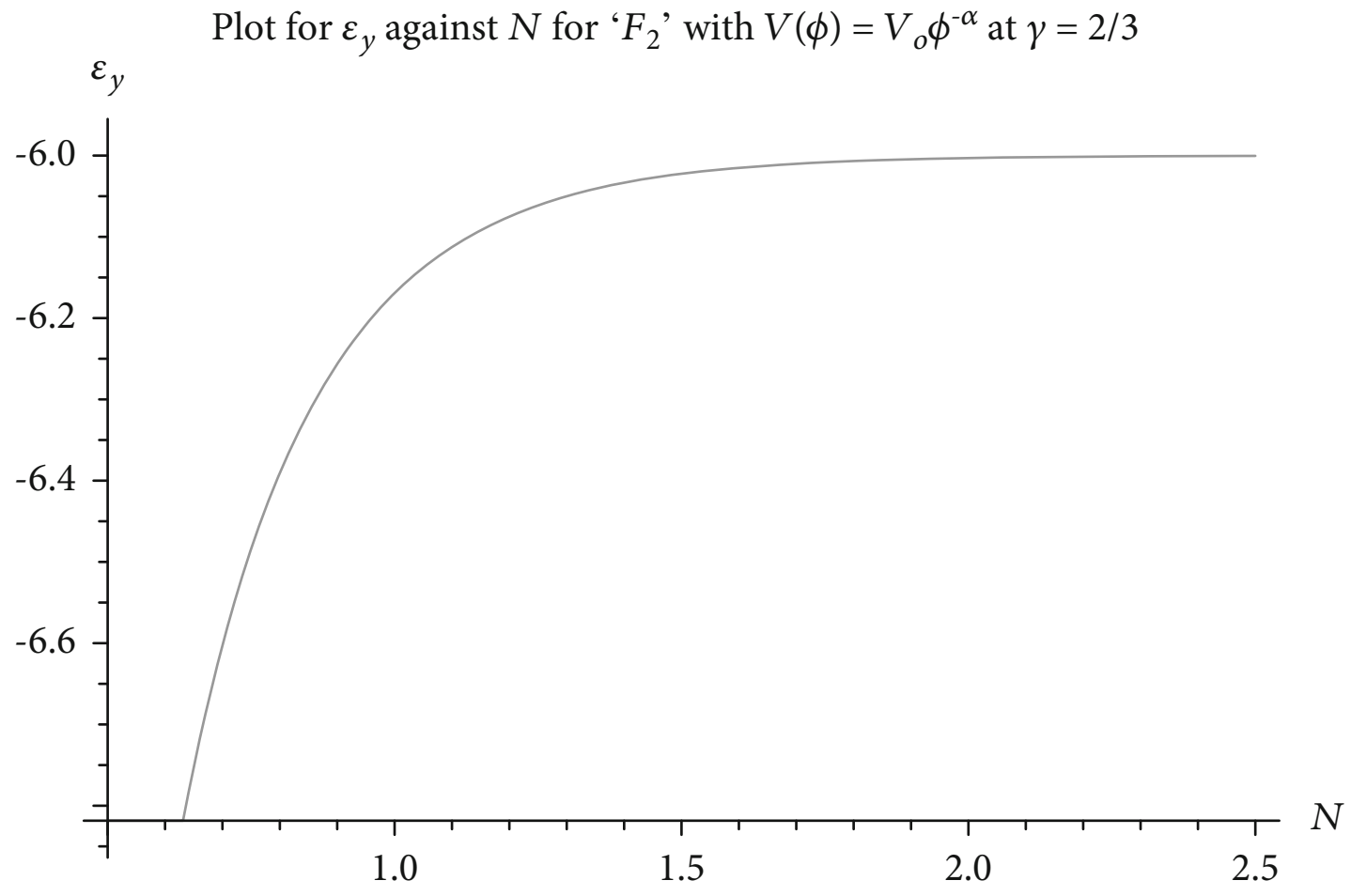

FIGURE 21: Shows the variation of perturbations along the $y$-axis ($\epsilon_y$) against $N$ at fixed point $G_2$ for the inverse power law potential form.

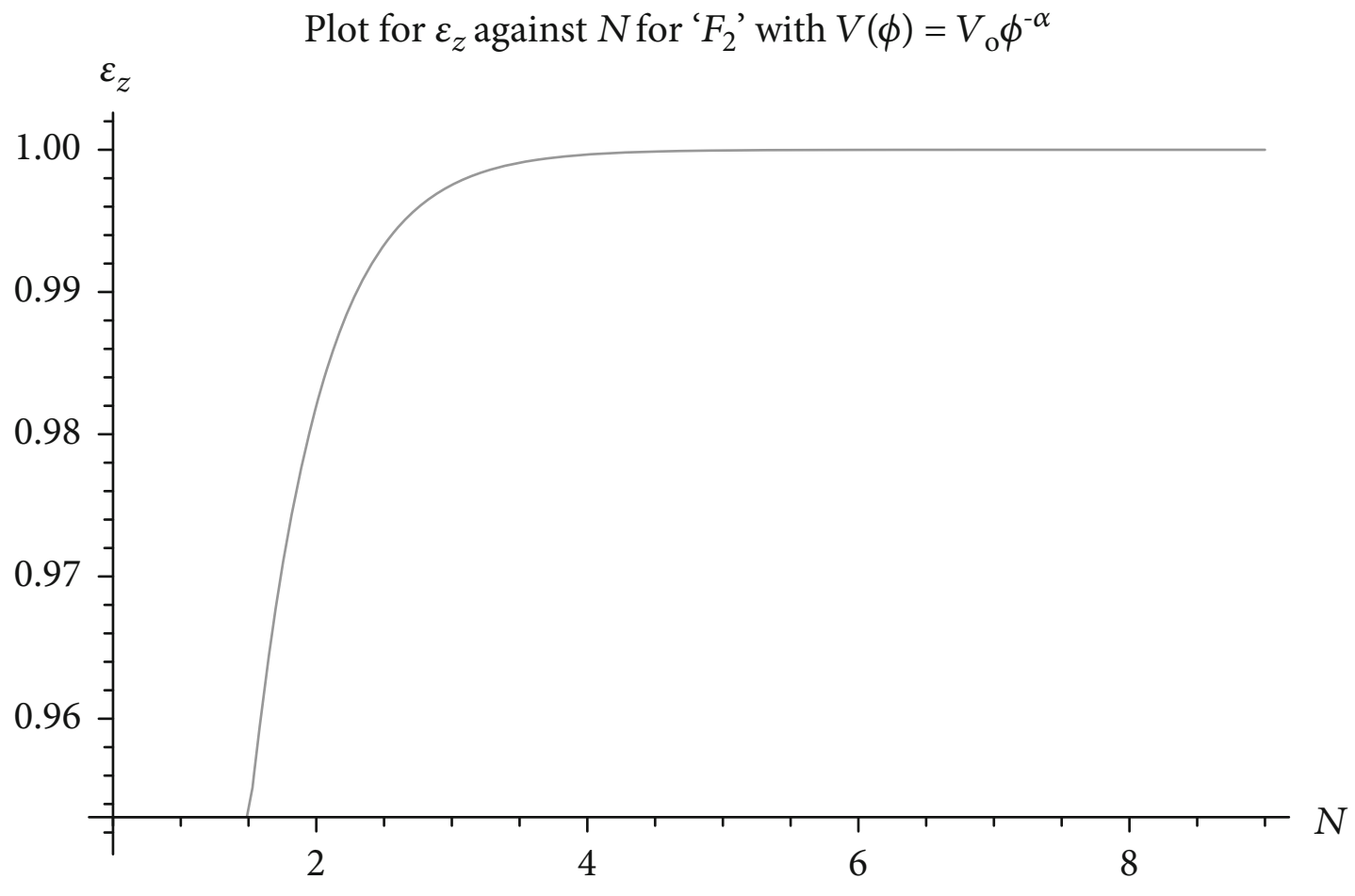

FIGURE 22: Shows the variation of perturbations along the $z$-axis ($\epsilon_z$) against $N$ at fixed point $G_2$ for the inverse power law potential form.

Table 4: Table for dynamical system with inverse power law potential form.

| Fixed points | $x$ | $y$ | $z$ | $\lambda$ | Eigenvalues | $\Omega_\phi$ | Stability | Behavior | $\omega_\phi$ |
|---|---|---|---|---|---|---|---|---|---|
| $G_1$ | 0 | 0 | 1 | 0 | $0, -2, -3, -3\gamma$ | 1 | Stable for all $\gamma \in [0, 2]$ | Stable, late time attractor | -1 (dark energy model) |
| $G_2$ | 1 | 0 | 0 | 0 | 4, 4, 6, 0 | 1 | Stable for all $\gamma \neq 0$ | Stable for $\gamma \in (0, 2]$ | 1 (ultrarelativistic fluid) |

$k = 0$. So our model gives a more generalized way of study by taking into account the curvature constant ($k$). We get three fixed points $A$, $B$, and $C$ tabulated in Table 2. By perturbing the system by a small amount ($\epsilon$), we find the perturbation along each axis as a function of $N$ where we have found $A$ to be stable for $\sqrt{2} < \lambda < \sqrt{3}$. We have also found the Jacobian matrix ($J$) and evaluated the eigenvalues through its characteristic polynomial equation. All the eigenvalues are found to be negative. We also solve the value of $\Omega_\phi$ at $A$ and find its value to be $\Omega_\phi = 1.007$ which is in agreement with the high precision data in the Planck collaboration in cosmological parameters [47]. We found an effective equation of state $\omega_\phi = 0$ at the fixed point $A$. So the fixed point $A$ characterizes our cosmological model with a matter-dominated Universe. The fixed point $B$ is also stable for potentials with $\lambda < \sqrt{2}$, $\gamma > 2/3$, and $\Omega_\phi \sim 1$. The variation of perturbation along the $x$-, $y$-, and $z$-axes against logarithmic time ($N$) is shown in Figures 13–15. The value of $\omega_\phi$ for the fixed point $B$ comes out to be $\omega_\phi = -1$. This fixed point is an interesting one as it indicates that the Universe is dominated by the potential energy of the scalar field that drives the expansion of the Universe with acceleration. The fixed point $C$ is completely unstable as the perturbation along the $z$-axis grows exponentially as $N$ tends to infinity with $\Omega_\phi = 1$. We have also studied the dynamical system with different forms of potential, i.e., double exponential potential and inverse power law potential. In both the analyses, we get a four-dimensional dynamical system [48, 49]. Using the double exponential potential form and inverse power law potential form, we obtain a four-dimensional system with fixed points $F_1$ and $F_2$. For the analysis with the double exponential potential form, the fixed point $F_1$ is stable with $\Omega_\phi = 1$ and $\omega = -1$, and it behaves as a late time attractor. The presence of this fixed point in the system represents the dark energy model ($\omega_\phi = -1$), and the other fixed point $F_2$ is unstable with a stiff fluid model ($\omega_\phi = 1$). With the inverse power law potential form, both fixed points $G_1$ and $G_2$ are stable representing a dark energy model ($\omega_\phi = -1$) and a stiff fluid ($\omega_\phi = 1$), respectively. In our present work, we have revealed the expanding scenario of the Universe where we have analyzed for suitable values of $\gamma$ and $\lambda$. Our developed scalar field cosmological model shows a deep connection with the cosmic acceleration phenomena thereby supporting the fact that our Universe is in the phase of accelerated expansion.

## Data Availability

No data were used to support this study.

## Conflicts of Interest

The authors declare that there is no conflict of interest regarding the publication of this paper.

## Acknowledgments

We would like to thank the DST, SERB, India grant ECR/2016/001485 for the financial support.